\documentclass[sigconf]{acmart}
\AtBeginDocument{%
  \providecommand\BibTeX{{%
    \normalfont B\kern-0.5em{\scshape i\kern-0.25em b}\kern-0.8em\TeX}}}

\copyrightyear{2025}
\acmYear{2025}
\setcopyright{cc}
\setcctype{by}
\acmConference[DIS '25]{Designing Interactive Systems Conference}{July 5--9,
2025}{Funchal, Portugal}

\acmBooktitle{Designing Interactive Systems Conference (DIS '25), July 5--9,
2025, Funchal, Portugal}\acmDOI{10.1145/3715336.3735740}
\acmISBN{979-8-4007-1485-6/2025/07}

\acmConference[DIS '25]{Designing Interactive Systems Conference}{July 05--09,
  2025}{Madeira, Portugal}

\newcommand{\system}{\textsc{ParaScope}}
\definecolor{aliceblue}{rgb}{0.91, 0.94, 0.97}

\definecolor{revc}{RGB}{0, 0, 0} %{108, 0, 255} %revision color
\newcommand{\rev}[1]{{\color{revc}#1}}

\definecolor{darkgreen}{RGB}{34,139,34}
\definecolor{pBlue}{HTML}{B8E3E9}
\definecolor{vPink}{HTML}{F8C8DC}
\definecolor{eLime}{HTML}{CCEFB9}

\newcommand{\flabel}[1]{\tcbox[on line, frame empty, boxsep=0pt, left=2pt, right=2pt, top=2pt, bottom=2pt, colback=pBlue]{#1}}
\newcommand{\glabel}[1]{\tcbox[on line, frame empty, boxsep=0pt, left=2pt, right=2pt, top=2pt, bottom=2pt, colback=eLime]{#1}}

\newcommand{\bglobal}{\flabel{\textsc{Global}}}
\newcommand{\blocal}{\flabel{\textsc{Local}}}

\newcommand{\fscore}{AI Score}
\newcommand{\ftranslation}{AI Translation}
\newcommand{\fexplanation}{AI Explanation}
\newcommand{\fexample}{Example Sentence}
\newcommand{\ffrequency}{Frequency}

\newcommand{\textft}[1]{{\fontfamily{lmss}\selectfont{#1}}}

\usepackage{booktabs}
\usepackage{xspace}
\usepackage{xcolor}
\usepackage{caption}
\usepackage{subcaption}
\usepackage{colortbl}
\usepackage{multirow}
\usepackage{graphicx}
\usepackage{framed}
\usepackage{csquotes}
\usepackage{lipsum}
\usepackage{enumitem}
\usepackage{framed}
\usepackage[most]{tcolorbox}
\usepackage{array}
\usepackage{makecell}

\makeatletter
\@ifpackageloaded{arydshln}{
   \newcommand{\dashedline}[1]{
    \cdashline{#1}[.4pt/1pt]\noalign{\vskip 1pt}
  }
}{
  \newcommand{\dashedline}[1]{
    \cmidrule[.0025mm]{#1}
  }
}
\makeatother

\begin{document}

\title[Design Opportunities for Explainable AI Paraphrasing Tools: A User Study with Non-native English Speakers]{Design Opportunities for Explainable AI Paraphrasing Tools: \\A User Study with Non-native English Speakers}

\author{Yewon Kim}
\affiliation{%
    \institution{KAIST}
    \city{Daejeon}
    \country{Republic of Korea}
}
\email{yewon.e.kim@kaist.ac.kr}

\author{Thanh-Long V. Le}
\affiliation{%
  \institution{KAIST}
  \city{Seoul}
  \country{Republic of Korea}}
\email{thanhlong0780@kaist.ac.kr}

\author{Donghwi Kim}
\authornote{Work done while at KAIST.}
\affiliation{%
  \institution{Samsung Electronics}
  \city{Seoul}
  \country{Republic of Korea}
}
\email{dh.tony.kim@samsung.com}

\author{Mina Lee}
\authornotemark[2]
\affiliation{%
  \institution{University of Chicago}
  \city{Chicago, Illinois}
  \country{United States}
}
\email{mnlee@uchicago.edu}

\author{Sung-Ju Lee}
\authornote{Equal senior role.}
\affiliation{%
  \institution{KAIST}
  \city{Daejeon}
  \country{Republic of Korea}
}
\email{profsj@kaist.ac.kr}

%%
%% By default, the full list of authors will be used in the page
%% headers. Often, this list is too long, and will overlap
%% other information printed in the page headers. This command allows
%% the author to define a more concise list
%% of authors' names for this purpose.
% \renewcommand{\shortauthors}{Kim et al.}

\begin{abstract}
\rev{
We investigate how non-native English speakers (NNESs) interact with diverse information aids to assess and select AI-generated paraphrases. We develop \system{}, an AI paraphrasing assistant that integrates diverse information aids, such as back-translation, explanations, and usage examples, and logs user interaction data. Our in-lab study with 22 NNESs reveals that user preferences for information aids vary by language proficiency, with workflows progressing from global to more detailed information. While back-translation was the most frequently used aid, it was not a decisive factor in suggestion acceptance; users combined multiple information aids to make informed decisions. Our findings demonstrate the potential of explainable AI paraphrasing tools to enhance NNESs’ confidence, autonomy, and writing efficiency, while also emphasizing the importance of thoughtful design to prevent information overload. Based on these findings, we offer design implications for explainable AI paraphrasing tools that support NNESs in making informed decisions when using AI writing systems.
}
\end{abstract} 

%%
%% The code below is generated by the tool at http://dl.acm.org/ccs.cfm.
%% Please copy and paste the code instead of the example below.
%%


%%
%% Keywords. The author(s) should pick words that accurately describe
%% the work being presented. Separate the keywords with commas.

\maketitle

\vspace{1em}
\section{Introduction}

\rev{
We have witnessed an enormous shift in the capability of AI writing tools from spell checkers~\cite{grammarly, languagetool} to content generators %capable of 
producing fluent, human-like text~\cite{10.1145/3613904.3642105, 10.1145/3579592, huang2023role}. 
These tools open new avenues for non-native English speakers (NNESs), who face unique linguistic and cultural barriers when writing in English~\cite{fareed2016esl, emailcommunication, amano2023manifold, worddecipher, 10.1371/journal.pbio.3002184, lim2022understanding}. 
For instance, NNES-authored emails may unintentionally include awkward or contextually inappropriate expressions, reducing their likelihood of receiving replies~\citep{lim2022understanding}. Such expressions may also negatively influence recipients' perceptions of the NNES sender's intelligence and trustworthiness, 
even when the recipient is aware of the sender's foreign background~\cite{emailcommunication}. 

Amid these growing challenges faced by NNESs, recent advances in large language models (LLMs)~\citep{gpt4o, instructgpt, touvron2023llama} have significantly expanded the scope of AI writing support. 
Equipped with capabilities to generate fluent, human-like text~\citep{10.1145/3591196.3596612, zhang-etal-2024-llm}, compared to traditional rule-based feedback tools~\citep{awkchecker, zhang-etal-2016-argrewrite}, LLMs now assist with a broad range of writing tasks, including idea brainstorming~\citep{10.1145/3563657.3595996, 10.1145/3532106.3533533, 10.1145/3613904.3642105}, drafting~\citep{coauthor, siddiqui2025script, dang2023choice}, and paraphrasing~\citep{langsmith, rewriting}. 
Among these myriad writing tasks that AI can support, the paraphrasing task is shown to improve NNESs' writing by suggesting more fluent paraphrases than their original sentences~\cite{chen2015developing}.}

% Research has been focusing on how to support these challenges that NNESs go through. traditionally by grammar checkers~\cite{awkchecker} and feedback tools~\cite{zhang-etal-2016-argrewrite} that correct errors based on predefined rules or statistical machine learning models.
% Recent advances in large language models (LLMs)~\citep{gpt4o, instructgpt, touvron2023llama} have significantly expanded the capabilities of these writing tools, enabling them to generate fluent, human-like text~\citep{10.1145/3591196.3596612, zhang-etal-2024-llm}. These advanced models support a wide range of writing tasks beyond rule-based feedback tools~\citep{awkchecker, zhang-etal-2016-argrewrite}, including idea brainstorming~\citep{10.1145/3563657.3595996, 10.1145/3532106.3533533, 10.1145/3613904.3642105}, drafting~\citep{coauthor, siddiqui2025script, dang2023choice}, and paraphrasing~\citep{langsmith, rewriting}, just to name a few. 
% Among these myriad writing tasks that AI can support, the paraphrasing task is shown to improve NNESs' writing by suggesting more fluent paraphrases than their original sentences~\cite{chen2015developing}.

While an AI paraphrasing tool could instantly offer paraphrased suggestions, the ultimate responsibility of assessing and selecting the most context-appropriate falls on users. This introduces a significant challenge for NNESs, as they may lack the linguistic proficiency and cultural awareness to evaluate these suggestions accurately~\citep{formative}. 
Prior research has identified several types of information aids, such as suggestion quality scores and example sentences, that can assist in evaluating AI-generated suggestions~\citep{langsmith, awkchecker, liu2000pens, napolitano2009techwriter}. It has recommended developing writing tools that integrate such aids to provide comprehensive support~\citep{rewriting}. 
\rev{
However, key questions remain open: 
\textit{How can information aids be effectively integrated into writing assistants to collectively explain AI-generated paraphrase suggestions?}
Answering this question necessitates a deeper understanding of how NNESs interact with and respond to information aids during paraphrasing tasks---an area that remains underexplored.
}

To address this gap, we investigate how NNESs interact with diverse information aids (interchangeably referred to as ``information aids'' and ``features'' throughout the paper) to assess and select paraphrase suggestions. 
\rev{
Concretely, we developed \system{}, a research prototype 
that integrates AI paraphrasing with five types of information aids within a single writing interface.
These aids---\fscore{}, \ftranslation{}, \fexplanation{}, \fexample{}, and \ffrequency{}---represent support information commonly used by NNESs in real-world writing contexts. \system{} logs user interactions with both paraphrase suggestions and the accompanying aids~(Section~\ref{section:prototype}).
Through a lab-based evaluation of \system{} with 22 NNESs writing academic emails~(Section~\ref{section:main}), we examined participants' feature usage patterns, perceptions, and written outcomes~(Section~\ref{section:findings}). 
}
We conclude with design implications for explainable AI paraphrasing tools tailored to NNESs (Section~\ref{section:discussion}).

Our findings reveal that while participants generally preferred features used to gauge the overall quality of the suggestions, e.g., \fscore{}, \ftranslation{}, and \fexplanation{}, over features offering detailed explanations of specific words or expressions like \fexample{} and \ffrequency{}, their preferences varied by proficiency level. 
NNESs with lower proficiency used significantly more information aids than those with higher proficiency and demonstrated a particular preference for \fscore{}, a numeric representation of suggestion quality.
Although \ftranslation{}---back-translating English suggestions into users' first languages---was the most frequently used feature, it was not a decisive factor in suggestion acceptance; users relied on a combination of information aids to make informed decisions. 
Post-interview findings further revealed that integrating information aids positively influenced perceived efficiency, confidence, and autonomy in the decision-making process, although participants noted the potential for information overload if features were integrated without consideration.

Through this study, we discuss design implications for creating explainable AI paraphrasing tools that support NNESs in making informed decisions when collaborating with AI systems. We also explore opportunities for explainable writing assistants in broader contexts, offering directions for future research to enhance writing support across diverse user needs.

We summarize our contributions as follows:
\begin{itemize}
    \item \rev{\system{}, a research artifact prototyping an AI paraphrasing support system that integrates a paraphrasing tool with five information aids to investigate and compare common strategies NNESs use for paraphrasing.}
    \item \rev{Empirical findings from an in-lab user study of \system{} with 22 NNESs in an academic email writing context, combining qualitative and quantitative data on how participants engaged with information aids to assess AI suggestions and how these aids shaped their user experience.}
    \item Design implications for explainable AI paraphrasing tools tailored to the needs of NNESs. 
\end{itemize}
\section{Related Work}

\subsection{NNESs' Challenges in Writing}

Writing is regarded as one of the most challenging and complex tasks among NNESs~\cite{Ariyanti2017/10}. NNESs' written English displays lower linguistic accuracy and complexity compared with native speakers~\cite{awkchecker, polio1997measures, ortega2003syntactic}. 
\rev{
Their texts tend to include more spelling and grammatical errors and exhibit less lexical diversity~\citep{ridley-etal-2023-addressing}.}
NNESs also face challenges in selecting contextually appropriate words or expressions, primarily due to the gaps in their cultural background~\cite{varonis1985miscommunication}. 
These issues become particularly salient in high-stakes written comunication scenarios, such as email and academic writing~\cite{emailcommunication, e44e8cc1-cf0e-3d0d-bd34-e07d16821d0b, jbp:/content/journals/10.1075/aila.20.04flo,HUANG201033, chatty}. 
In emails, linguistic errors can lead recipients to NNES senders as less intelligent, diligent, conscientious, and cognitively trustworthy~\cite{emailcommunication}, and NNESs' overuse of casual language in academic writings can leave them at a competitive disadvantage~\cite{chatty}. 
\rev{
Furthermore, NNESs’ writing is often misclassified as AI-generated by automated detectors due to its lower lexical complexity, raising concerns around fairness~\citep{LIANG2023100779}.}

\subsection{AI Writing Support Tools for NNESs}
While AI writing support tools span a broad range of tasks ranging from idea brainstorming~\cite{elephant, lyrisys, sparks, talebrush, characterchat, 10.1145/3563657.3595996, 10.1145/3532106.3533533, 10.1145/3613904.3642105} to revision~\cite{thiemo22empathy, 10.1145/3411764.3445683, 10.1145/3532106.3533526}, those designed for NNESs often focus on the revision process of writing. These tools aim to aid improvement in writing quality by providing assessment and feedback to the NNESs' writing, primarily through grammar and spelling checkers~\cite{awkchecker, rei-yannakoudakis-2016-compositional, gamon2010using, liu2000pens, chang2015writeahead2} and feedback generation tools~\cite{burstein2003toward, 10.1145/3272973.3274069, hanawa-etal-2021-exploring, nagata2019toward, kinnunen2012swan}. Another way to assist NNESs' revision process is paraphrasing~\cite{langsmith, rewriting, mita2022automated, jiang-etal-2022-arXivEdits, du-etal-2022-understanding-iterative, yimam-biemann-2018-demonstrating}, which suggests alternative ways of expressing the same content, potentially enhancing the quality of original texts written by NNESs~\cite{langsmith}.

Despite the usefulness of revision tools, NNESs' lack of linguistic proficiency hinders them from accurately assessing and selecting tools' suggestions~\cite{rewriting, formative, KNUTSSON20071122}. As a solution, several revision tools provide information aids (e.g., dictionary results, grammar rules) along with suggestions to help users better understand and reason about the suggestions~\cite{liu2000pens, awkchecker, chang2015writeahead2, langsmith, rewriting}. For instance, lexical and grammatical suggestions are provided along with informational aids such as dictionary samples and grammar patterns sourced from an academic corpus~\cite{chang2015writeahead2}. On the other hand, LangSmith, a paraphrasing tool for NNES~\cite{langsmith, rewriting}, provides a ``typicality score'' of paraphrased suggestions, representing the suggestion qualities. Nevertheless, we do not clearly understand which and how supports should be provided~\cite{rewriting} to best support NNESs using AI paraphrasing tools. We aim to bridge this gap, offering insights into effective support strategies for NNESs engaged in paraphrasing activities.

\subsection{Empirical Studies of Writing with AI}
Previous research has explored user behaviors and perceptions of AI writing tools. A major thread of this research has focused on predictive text systems, which aid users in composing text by suggesting the next phrases or sentences they might use~\cite{arnold2016suggesting, multipleparallel, coauthor, dang2023choice, 10.1145/3544548.3581351}. Recent work~\cite{10.1145/3544548.3581351} explored how users write email replies with sentence vs. message-level suggestions. Another study~\cite{multipleparallel} analyzed differences between native and non-native English speakers concerning their perception of and writing behaviors influenced by writing tools' next phrase suggestions. 
\rev{In parallel, a line of research studied user behaviors by providing AI-based analytics to user written text, such as event sequence visualizations and paragraph summaries~\citep{chen2025comparing, 10.1145/3526113.3545672, shibani2023visual}. 
For instance, COALA~\citep{chen2025comparing} compared native and non-native English speakers' collaborative writing behaviors when they are equipped with visual analytics.}
On the other hand, a few studies have investigated user behaviors regarding paraphrasing tools~\cite{rewriting, quillbotutilization}. 
Closest to our study, researchers explored behaviors of NNESs with Langsmith~\cite{langsmith} and found that NNESs use external resources such as translators and web search results when paraphrasing with AI~\cite{rewriting}. However, they focused on the impact of translator use on NNESs' engagement with the tool's outputs. Our study takes a broader and more detailed approach by identifying a comprehensive set of information aids that NNESs need when paraphrasing with AI and comprehensively analyzing user behaviors given these aids. Moreover, we provide insights on designing paraphrasing tools with information aids to support NNESs effectively.
\section{\system{} Design and Implementation}
\label{section:prototype}

\subsection{\system{} Design Principles}
\rev{
We designed \system{} as a research prototype to investigate how NNESs interact with diverse information aids when assessing and selecting AI-generated paraphrases. 
To achieve this, we established high-level design principles of \system{} drawing on prior work~\citep{rewriting, gamut, 10.1145/3544548.3581351, 10.1145/642611.642616, 10.1145/3544548.3581388, 10.1145/3581641.3584054, privateyetsocial, coauthor}. Our design aimed to:
}

% \system{} is designed as a technology probe to explore how NNESs interact with information aids in AI-assisted writing, to inform the design of future AI writing support systems tailored to their needs. 
% A technology probe is ``an instrument deployed to discover the unknown--to hopefully return with useful or interesting data''~\citep{10.1145/642611.642616}. 
% Ideally, technology probes should achieve three disciplinary goals: (i) \textit{engineering goal}: installing technology into a real-world context; (ii) \textit{design goal}: enabling simple, flexible, and open-ended use to support users' needs and inspire ideas for new technologies; (iii) \textit{social science goal}: collecting information about the needs and desires of users in-situ~\citep{10.1145/642611.642616}. 
% Technology probes are widely used in human-computer interaction research as a contextual and participatory method~\citep{gamut, privateyetsocial, 10.1145/291224.291235, 10.1145/3544548.3581388, 10.1145/3581641.3584054, 10.1145/3064663.3064715}.
% In designing \system{}, we formulated the following high-level goals:

\begin{itemize}[leftmargin=1.5em]
    \item \glabel{DP1}\rev{\textbf{Include core strategies NNESs employ as information aids in a realistic writing context}: 
    Prior studies demonstrate that realistic task contexts elicit natural user behaviors~\citep{rewriting, 10.1145/3544548.3581351, gamut,  10.1145/642611.642616}.
    We aim to develop an AI-assisted paraphrasing tool that integrates multiple types of information aids commonly used by NNESs, and embed it in a writing task representative of real-world scenarios faced by NNES users. 

    \item \glabel{DP2}\rev{\textbf{Support open‑ended, user‑driven exploration}: 
    Prior studies show that minimally constrained interaction reveal emergent usage strategies and inspire design directions~\citep{gamut, 10.1145/3544548.3581388, 10.1145/3581641.3584054, privateyetsocial}.
    We focus on designing a system that allows participants to freely engage with different information aids during the paraphrasing process in an open-ended manner.}
    }
    
    \item \glabel{DP3}\rev{\textbf{Capture user-system interaction data}: 
    Prior HCI work demonstrates that recording user interactions with a system provides insights into user behavior with interactive systems and reveals patterns that inform future system design \citep{coauthor, privateyetsocial, 10.1145/642611.642616}. 
    We log detailed user interactions and collect qualitative feedback to understand how participants engage with the system. }

\end{itemize}

\subsection{Overall Design of \system{}}
\begin{figure}
    \centering
    \includegraphics[width=\linewidth]{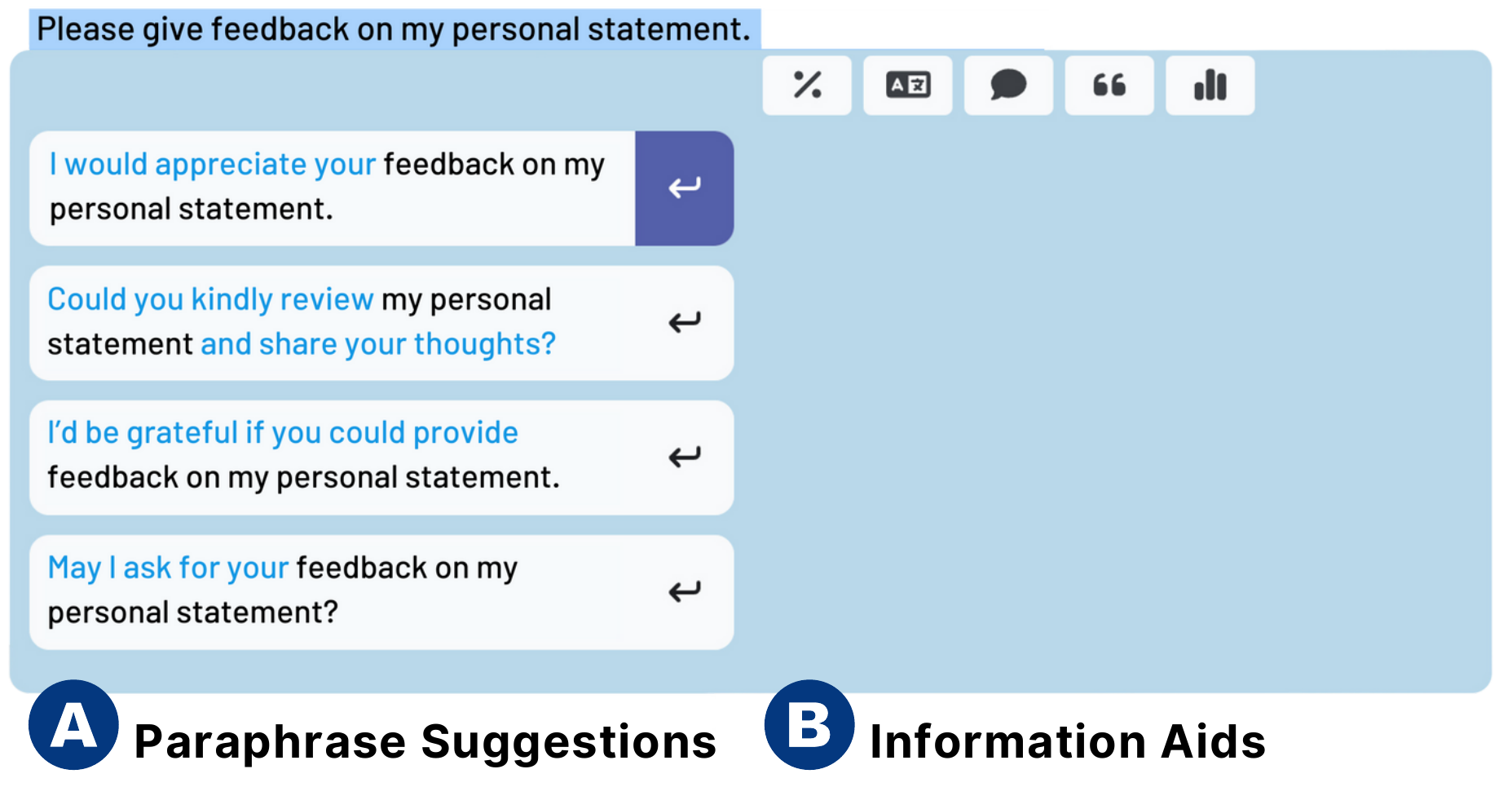}
    \vspace{-1em}
    \caption{Overview of the \system{} interface. The interface features two panes: (A) a pane displaying four paraphrased suggestions for the user’s input text and (B) a pane providing information aids. The right pane (B) remains empty until the user interacts with an information aids button.}
    \vspace{-1.5em}
    \Description{This figure shows the overview of the system's interface. The left panel displays four paraphrased suggestions for the user's input text and the right panel provides five information aids buttons.}
    \label{fig:prototype:overview}
\end{figure}

\system{} is an AI writing support system that integrates paraphrasing functionality with diverse information aids in a unified interface, 
enabling NNESs to make informed decisions when assessing and selecting AI paraphrased suggestions.
The system consists of two main components: the paraphrasing pane (Figure~\ref{fig:prototype:overview}(A)) and the information aids pane (Figure~\ref{fig:prototype:overview}(B)).
In the paraphrasing pane, users can navigate and select one of four paraphrased suggestions for their input text.
In the information aids pane, users can flexibly explore five types of information aids: \fscore{}, \ftranslation{}, \fexplanation{}, \fexample{}, and \ffrequency{} to evaluate the system's suggestions. 

\subsubsection{Paraphrasing in the email writing task as a use case \textup{(\glabel{DP1})}} 
To contextualize \system{} in a meaningful real-world setting, we selected the task of paraphrasing in email writing.
Writing emails in English is a common yet challenging activity for NNESs, as it requires clear, professional communication that adheres to linguistic and cultural norms they may be unfamiliar with~\citep{emailcommunication}.
Moreover, paraphrasing---an essential activity for refining tone, improving the wording, and enhancing clarity---is widely practiced by NNESs in both practical and educational contexts, often with the aid of AI writing support systems~\citep{rewriting, formative}.
The practical relevance of paraphrasing and email writing makes this task a suitable context for studying user interactions with AI-generated suggestions and information aids, providing insights to inform the design of writing support systems for NNES users.

\subsubsection{UI design and interaction paradigm of \system{} \textup{(\glabel{DP2})}}
We aim to understand how NNESs engage with information aids when assessing and selecting AI-generated suggestions. To achieve this, we focused on two key aspects in designing \system{}:

\begin{table*}[t]
  \centering
  \caption{List of logged events. Each event is represented as a tuple containing the event name, timestamp, and snapshot of the editor. Text events have associated metadata containing information on inserted or deleted text. Cursor events have associated metadata containing information on start and end indices of cursor selection.}
  \resizebox{\textwidth}{!}{
  \begin{tabular}{p{2.5cm}p{11cm}p{4cm}}
    \toprule
    Event name & Trigger & Description \\
    \midrule
    \multicolumn{3}{l}{\cellcolor{gray!20}Category: Session} \\
    \addlinespace[2mm]
    \textft{session-start} & System initializes the editor &Start writing session\\
    \textft{session-end} & User clicks submit button &Finish writing session \\
    \addlinespace[1mm]
    \hline
    \addlinespace[1mm]
    \multicolumn{3}{l}{\cellcolor{gray!20}Category: Text} \\
    \addlinespace[2mm]
    \textft{text-insert} & User/system inputs (any key) & Insert text \\
    \textft{text-delete} & User presses \texttt{delete} key & Delete text \\
    \addlinespace[1mm]
    \hline
    \addlinespace[1mm]
    \multicolumn{3}{l}{\cellcolor{gray!20}Category: Cursor} \\
    \addlinespace[2mm]
    \textft{cursor-forward} & User/system inputs \{$\downarrow$, $\rightarrow$\} key & Move cursor forward \\
    \textft{cursor-backward} & User presses \{$\uparrow$, $\leftarrow$\} key & Move cursor backward \\
    \textft{cursor-select} & User presses \texttt{shift} + \{$\downarrow$, $\rightarrow$, $\uparrow$, $\leftarrow$\} & Select range of text \\
    \addlinespace[1mm]
    \hline
    \addlinespace[1mm]
    \multicolumn{3}{l}{\cellcolor{gray!20}Category: Suggestion} \\
    \addlinespace[2mm]
    \textft{suggestion-get} & User presses \texttt{cmd}/\texttt{ctrl} + \texttt{j} & Request new suggestions \\
    \textft{suggestion-open} & System fetches suggestions & Show interface w/ suggestions \\
    \textft{suggestion-reopen} & User presses \texttt{shift} + \texttt{tab} & Reopen interface w/ prev. suggestions \\
    \addlinespace[-0.5mm]
    \multicolumn{3}{l}{\emph{While interface is displayed}} \\
    \addlinespace[1.5mm]
    \textft{suggestion-select} & User presses \texttt{enter} + suggestion is focused & Select suggestion \\
    & User clicks a suggestion & \\
    \addlinespace[1mm]
    \textft{suggestion-close} & User/system inputs \texttt{esc} or (any key) & Hide interface \\
    & User clicks outside of interface & \\
    \addlinespace[1mm]
    \hline
    \addlinespace[1mm]
    \multicolumn{3}{l}{\cellcolor{gray!20}Category: Information Aid} \\
    \addlinespace[2mm]
    \multicolumn{3}{l}{\emph{While interface is displayed}} \\
    \addlinespace[2mm]
    \textft{info-get} & User presses \texttt{enter} + information aid button is focused & Request information aid \\
    & User clicks the information aid button & \\
    \addlinespace[1mm]
    \textft{info-open} & System fetches queried information aid & Show requested information aid \\
    \addlinespace[1mm]
    \textft{info-expl-select} & User presses \texttt{enter} + suggestion is focused + \fexplanation{} is opened & View explanation associated \\
    & User clicks suggestion + \fexplanation{} is opened & with suggestion \\
    \addlinespace[1mm]
    \textft{info-exam-query} & User presses \texttt{enter} + \fexample{} is opened + example search bar is focused & Query example sentences \\
    & User clicks the example search button + \fexample{} is opened & \\
    \addlinespace[1mm]
    \textft{info-freq-query} & User presses \texttt{enter} + \ffrequency{} is opened + frequency search bar is focused & Query frequencies \\
    & User clicks the frequency search button + \ffrequency{} is opened & \\
    \arrayrulecolor{black}\bottomrule
  \end{tabular}
  }
  
  \Description{The table lists the type of log events, which consists of event name, trigger, and the description of the event.}
  \label{tab:log_events}
\end{table*}

\paragraph{Usability consistency}
Since our primary focus is on studying the use of information aids, we ensured usability consistency~\citep{designspace}--alignment of user experience with other systems that the user is familiar with--in designing the surrounding elements, i.e., the overall UI and interaction methods, of the system.
This approach reduces the cognitive overhead associated with adapting to unfamiliar systems, allowing participants to focus on exploring and utilizing the information aids. 
Specifically, we drew on conventions established by existing writing assistants~\citep{wordtune, writefull, langsmith, grammarly, coauthor}, which provide real-time support directly within text editors.
In \system{}, we implemented a pop-up interface activated via a keyboard shortcut. 
When a user selects a range of text\footnote{The system automatically adjusts the user's selection to encompass complete words if the initial selection ends mid-word.} and presses \texttt{cmd} or \texttt{ctrl} + \texttt{j}, a pop-up box appears directly below the cursor (Figure~\ref{fig:prototype:overview}). 
The system supports navigation of suggestions and information aids using both a mouse (point and click) and a keyboard (arrow keys to navigate UI components, \texttt{Enter} to accept a suggestion or open an information aid).
Additionally, we provided multiple paraphrased suggestions, aligning with established practices in AI paraphrasing tools~\citep{wordtune, langsmith, rewriting, 10.1145/2505984}. We chose the number of suggestions as four, based on the prior work~\citep{multipleparallel} which demonstrated NNESs benefit most from receiving 3<$N$<6 parallel suggestions.

\paragraph{Flexible, open-ended exploration of information aids}
To promote open-ended exploration of information aids, \system{} displays the features---\fscore{}, \ftranslation{}, \fexplanation{}, \fexample{}, and \ffrequency{}---as horizontally aligned buttons. 
The UI does not select any feature independently, and the right pane remains empty until a user clicks any button, ensuring all features are equally accessible and users flexibly explore suggestions based on their needs. 
We provide further details on the design and functionality of each information aid in \S~\ref{section:system:feature}.

\begin{figure*}
    \centering
    \includegraphics[width=0.95\linewidth]{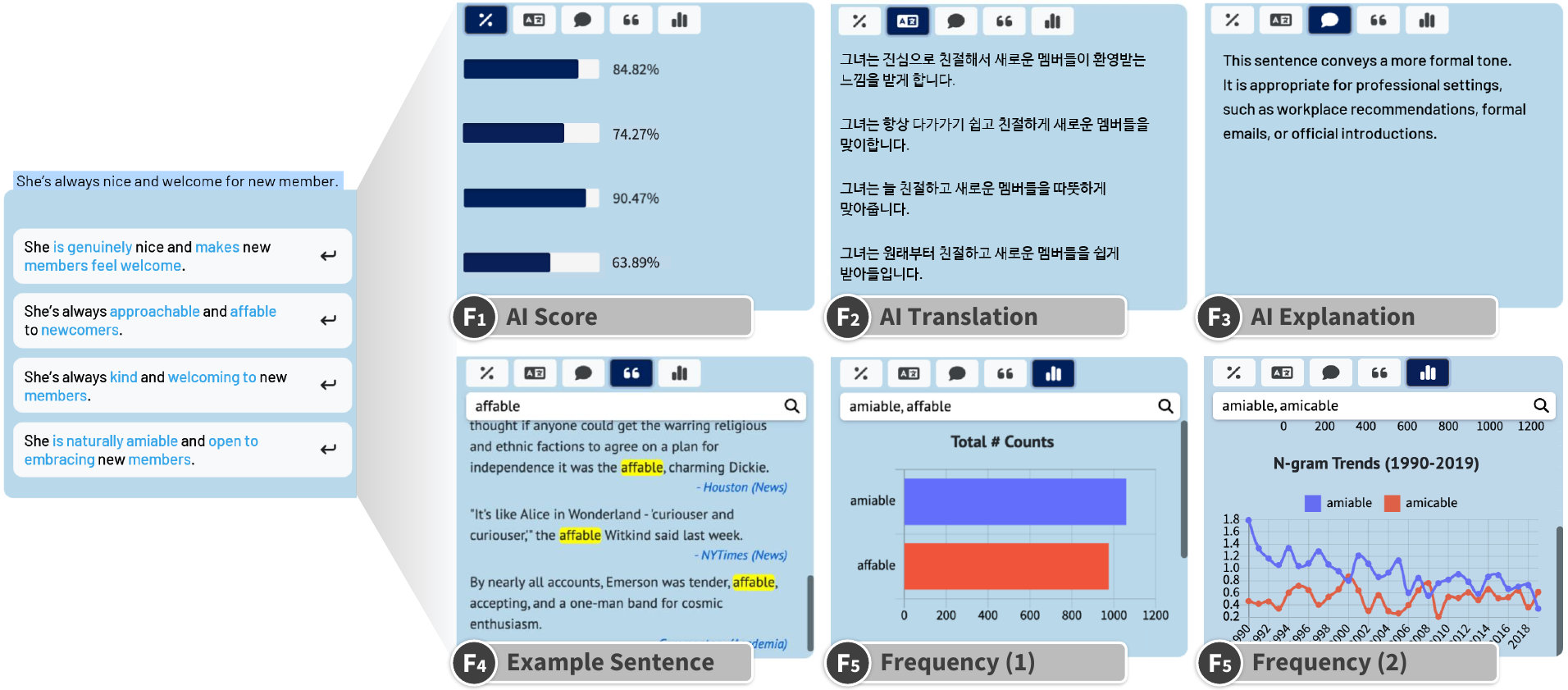}
    \caption{Illustrations of the information aids: \fscore{} ($F_1$), \ftranslation{} ($F_2$), \fexplanation{} ($F_3$), \fexample{} ($F_4$), and \ffrequency{} ($F_5$) for the four paraphrased suggestions of the original text (``She's always nice and welcome for new member.''). 
    $F_1$ (\fscore{}): The scores representing paraphrased suggestions' quality are displayed next to each suggestion as both bar graphs and numerical values. $F_2$ (\ftranslation{}): Translated versions of suggestions in users' first language (in this figure, Korean) are displayed next to each suggestion. $F_3$ (\fexplanation{}): Textual explanation about tones and appropriate contexts for each suggestion is displayed. $F_4$ (\fexample{}): After users search a term using the search box, example sentences containing the term are displayed, allowing users to browse the sentences. $F_5$ (\ffrequency{}): After users search a term(s) using the search box, a bar graph visualizing the frequency of the term(s) is displayed. When users scroll down the prototype, they see a line graph depicting the number of occurrences of the term(s) over the years.}
    \vspace{-.5em}
    \Description{
    This figure illustrates the five support features: AI Score, AI Translation, AI Explanation, Example Sentence, and Frequency. AI Score shows the score of the sentences in percentage, along with an accompanying horizontal bar to express the score. AI Translation shows the paraphrased suggestion translated to the user's first language, which in this figure is Korean. AI Explanation shows a two-sentence explanation for the paraphrased suggestion. Example Sentence shows the queried phrase "affable" inside the search box and the search results. Retrieved example sentences are shown with the matched phrase highlighted in yellow, and the source title and genre are on the far right. Frequency shows two queried phrases (amiable, affable) inside the search box, and the search results - two horizontal bars with different colors expressing the total occurrence of those phrases. When a user scrolls down the Frequency pane, it shows the line graph depicting the frequency trend of the search queries from 1990 to 2019.
    }
    \label{fig:prototype:features}
\end{figure*}

\subsubsection{Logging user-system interactions \textup{(\glabel{DP3})}}
To understand how users interact with information aids to assess and select paraphrased suggestions, we designed a logging mechanism to capture user-system interaction data. 
Inspired by the prior work~\citep{coauthor}, we record user-system interactions as events, represented as a tuple containing the event name, timestamp, and snapshot of the current editor and \system{}. 
An event can be inserting or deleting text within the editor, moving a cursor forward or backward, getting and navigating suggestions and information aids from the system, or accepting or dismissing suggestions. 
A complete list of all logged events is shown in Table~\ref{tab:log_events}. 
These logs provide granular details about user-system interactions, enabling the analysis of quantitative usage patterns, such as the preferences for specific information aids and the sequence of information aid usage.

%%%
\subsection{Design of Information Aids in \system{}}
\label{section:system:feature}
To design information aids that could be useful in evaluating AI-generated suggestions, we adopted a two-fold approach informed by prior research~\citep{gamut}. 
First, we identified information aids commonly used by NNESs by analyzing functionalities of existing writing support systems from both commercial tools~\citep{grammarly, wordtune, languagetool, ludwig, papago, quillbot, thesaurus, writefull, deepl, chatgpt, googletranslate, ngram, gdocautocorrect} and academic studies~\citep{langsmith, zhang-etal-2016-argrewrite, cahill-etal-2021-supporting, kinnunen2012swan, awkchecker, 10.1145/3272973.3274069, liu2000pens, napolitano2009techwriter, 10.1145/2505984}.
% For commercial tools, we conducted a survey with 141 NNESs recruited through university communities and social media to identify widely used writing tools. 
% For academic tools, we reviewed literature using terms such as ``writing tools'' and ``non-native speakers.''
Second, we conducted a 1-hour online formative study with 15 NNESs who reported regularly writing English emails and using AI writing assistants~\citep{grammarly, wordtune, quillbot, writefull, googletranslate} to gather information aids a user should be able to use when paraphrasing with AI. The participants' first language included Korean, Chinese, French, Filipino, and Hindi. 
All studies were conducted in a semi-structured manner online via Zoom. During the formative study, participants were asked to write a professional email in their naturalistic settings using AI writing assistants or online searches to aid their writing process. 
After the writing session, the researchers asked participants about their writing experiences, focusing on the tools and functionalities they use to overcome the difficulty of assessing and selecting AI-generated suggestions, as well as their suggestions for improving the writing process with AI tools. Participants were compensated with approximately 24 USD (35,000 KRW) for participants. The researchers transcribed all audio and conducted open coding and thematic analysis.

Together, we discussed and synthesized our findings into the following list of information aid features that a paraphrasing tool for NNESs should support, until all researchers reached a consensus. Specifically, we identified from the formative study that the major needs of NNESs in assessing and selecting AI-generated suggestions are two-fold: (i) gauging the overall quality of suggestions, and (ii) validating the real-world usage of specific words or expressions that appeared in the suggestions, both stemming from concerns about whether AI-generated texts they select might sound awkward. We analyzed and categorized tools or desired features mentioned by the formative study participants to address each need, and mapped each feature to the system features identified from the survey and literature review. 
This way, we operationalized user needs into concrete design features, ensuring relevance and applicability to the real world.
Accordingly, for each feature, we label it as \bglobal{} or \blocal{}, where \bglobal{} refers to features used to gauge the overall quality of the suggestion, and \blocal{} refers to features examining the usage of specific words or expressions in suggestions.

\subsubsection{\fscore{}  \textup{(\bglobal{})}}
AI-based \textit{numerical assessments} of suggestion quality provide users with metrics for evaluating AI-generated text, e.g., scores indicating suggestion qualities~\citep{langsmith, writefull, grammarly}. 
% For instance, LangSmith~\citep{langsmith} offers typicality scores alongside rewritten suggestions, which represent the suggestion qualities. Similarly, Writefull~\citep{writefull} provides probability scores of AI's paraphrased suggestions, which indicates the likelihood of each suggestion being accepted by other users. 
Formative study participants expressed similar needs, as noted by two participants. Notably, P8 stopped using Wordtune due to a lack of quality indicators, stating, \textit{``There are so many suggestions, but I have no clue which one is the winner,''} and suggested incorporating a \textit{``ranking system based on model confidence scores.''}

Motivated by these observations, we designed \fscore{}, which represents the quality of AI-generated paraphrases ($F_1$ in Figure~\ref{fig:prototype:features}).
We follow the convention of the existing tools~\citep{langsmith, writefull} to visualize the computed scores with horizontal bar graphs and numerical values in a percentage alongside AI suggestions.
To calculate the scores of paraphrased sentences, we aimed to quantify the following criteria: whether the suggestion is semantically similar to the original sentence, whether it presents diverse alternative expressions to the original text while ensuring syntactic accuracy, as defined by prior works in linguistics~\citep{chen-dolan-2011-collecting, 10.1162/COLI_a_00166}. 
For this purpose, we leveraged ParaScore~\cite{parascore}, a language model-based paraphrase evaluation metric. ParaScore outputs higher scores for paraphrases that (i) maintain a similar meaning to the original text 
and (ii) consist of diverse words and syntactic structures compared with the input, which aligns with our predefined goals.

\subsubsection{\ftranslation{} \textup{(\bglobal{})}} 
Back-translation is widely recognized in the literature as a critical strategy for NNESs to verify their comprehension of English sentences~\citep{Kobayashi1992EffectsOF, chamot1987study}. 
Participants in our survey reflected this practice, reporting various translation tools~\citep{languagetool, papago, deepl, googletranslate}.
Previous studies~\citep{liu2000pens, napolitano2009techwriter, 10.1145/2505984} have similarly incorporated back-translation in writing tools to enhance comprehension of English suggestions for NNESs.
Our formative study revealed similar patterns, with five participants highlighting back-translation as a key step for verifying AI-generated suggestions. For instance, P2 mentioned: \textit{``I use Naver Papago--I copy and paste all AI suggestions into it, read texts in my first language, and confirm whether the suggestions sound natural in English or not.''}

Based on these observations, we designed \ftranslation{}, a feature that translates paraphrased suggestions into a user's first language\footnote{As all participants in our user study (\S\ref{section:main:participants}) were native Korean speakers, translations were provided in Korean.} ($F_2$ in Figure~\ref{fig:prototype:features}). 
Drawing on prior work~\citep{10.1145/2505984, liu2000pens}, we displayed L1 translations alongside AI-generated suggestions to enable quick and efficient comparison between options.
For implementation, we utilized the Naver Papago API~\citep{papago}, a translator frequently mentioned by participants in the survey.

\subsubsection{\fexplanation{} \textup{(\bglobal{})}}
We identified the use of natural language explanations in writing support systems, either as a method for interpreting target English text~\citep{ludwig} or providing writing feedback~\citep{10.1145/3272973.3274069, kinnunen2012swan, cahill-etal-2021-supporting}.
For example, Ludwig~\citep{ludwig} provides explanations by clarifying whether a word or phrase is suitable for written English and detailing its usage contexts (e.g., ``It is typically used to express a warm greeting to someone who has joined a group, company, team, etc.''). 
Such explanations are also widely applied in other domains to enhance the interpretability of intelligent interactive agents~\citep{mariotti-etal-2020-towards, CAMBRIA2023103111, 8015489, 10.1007/978-3-642-04391-8_33, sokol2018conversational}.
% Notably, natural language explanations have been shown to improve decision-making under uncertainty~\citep{8015489} and accommodate a broader range of users by mimicking how humans typically explain their decisions verbally~\citep{10.1007/978-3-642-04391-8_33, sokol2018conversational}. 
Similarly, during our formative interviews, four participants sought natural language explanations on differences between paraphrases and original text, focusing on tone and appropriate usage contexts. 
P9 expressed a need for clarifying the tone of suggestions: \textit{``I wish the tool could tell me, `this option is more polite than the other.'’’} 
P5 also mentioned identifying suitable usage contexts: \textit{``It would be great if the tool informs me, `this option is better for academic writing.'’’}

Building on the findings, we designed \fexplanation{} with two goals: generate natural language explanations that (i) compare original and paraphrased text, and (ii) inform users about the tone and appropriate contexts of each suggestion ($F_3$ in Figure~\ref{fig:prototype:features}).
% For each paraphrased suggestion, a user can read a natural language explanation about how it differs from the original text in terms of tone and usage context. 
When users click the \fexplanation{} button, a natural language explanation for the first paraphrased suggestion appears in the right plan. 
Users can explore explanations for individual suggestions by selecting one at a time, ensuring that only one explanation is displayed at a time to reduce the cognitive load from presenting lengthy texts in a confined space~\citep{ally2005using}.
We implemented \fexplanation{} using OpenAI's GPT-3.5~\cite{gpt3.5turbo} with a few-shot prompting method~\cite{brownfewshot} (see Appendix~\ref{appendix:technical:explanation} for the prompt). 
GPT-3.5 was selected as it was the best available model at the time of the study and demonstrated strong capabilities in generating human-like responses~\citep{christiano2023deepreinforcementlearninghuman} and linguistic comprehension in few-shot settings~\citep{junjie2023comprehensive}.

\subsubsection{\fexample{} \textup{(\blocal{})}}
Participants from the survey often mentioned tools that provide example sentences for the search query~\citep{ludwig, thesaurus}. 
Related works also feature example sentences~\citep{liu2000pens, awkchecker} along with suggestions. 
For instance, AwkChecker provides example sentences of each suggestion to help users examine the context surrounding phrases to make an informed decision~\citep{awkchecker}.
Similarly, eight participants from the formative study referred to human-authored example sentences from credible sources like news, academic articles, or famous magazines to verify whether words or expressions are used in the real world. 
P11 mentioned checking example sentences from a native English corpus to verify AI suggestions: \textit{``I often refer to the Longman dictionary because it provides accurate information on how words are used in specific contexts. Its examples are taken from the native English corpus, so you can be confident that the words are used in the way that native English speakers use them.’’} 

We design \fexample{} to retrieve example sentences from credible sources (newspapers, academic journals, and magazines) containing a term specified by a user ($F_4$ in Figure~\ref{fig:prototype:features}). 
Initially, the search box appears in the right pane of the pop-up box, allowing the user to input a word or expression. Upon submitting it by clicking the search button (magnifier icon) or pressing the \texttt{enter} key, the tool retrieves sentences containing the input term (highlighted in yellow), randomly selects up to 10 example sentences, and shows them along with the reference and genre of the sentences (e.g., New York Times (News)). 
For the data source, we use the Corpus of Contemporary American English (COCA)~\cite{coca}. COCA is an extensive collection of over one billion words extracted from approximately 500,000 texts of diverse genres from 1990 to 2019. 
To ensure we retrieve only the sentences from credible sources, we selected only the sentences from three genres: academic journals, magazines, and newspapers.

\subsubsection{\ffrequency{} \textup{(\blocal{})}}
NNESs often verify whether a phrase is commonly used by analyzing the number of results returned by search engines~\citep{Guo2007BuildingAC, yi-etal-2008-web}. 
Similarly, several writing tools incorporate real-world word usage frequencies into their suggestions~\citep{10.1145/2505984, awkchecker, ngram}. 
For example, TransAhead provides grammatical patterns linked to user input and their frequencies in a reference corpus~\citep{10.1145/2505984}.
Three participants in our formative study also reported using statistical evidence to assess term usage. 
P11 described using Google search results to evaluate the prevalence of an expression: \textit{``One way to find out if native speakers commonly use an expression is to search it on Google and see how many results you get.’’} Similarly, P8 used Google Ngram Viewer~\citep{ngram} to check if the word or expression was trendy.

Inspired by tools that display total frequencies~\citep{10.1145/2505984, awkchecker} and visualize usage trends~\citep{ngram}, we designed \ffrequency{} to provide two types of information for user queries: total frequency counts and usage trends over time. 
After a user queries a term(s) using the search box, the right pane of the system displays (i)~a bar graph visualizing the total number of times each term appears in the source data and (ii)~a line graph depicting its frequency trend over the years (1990-2019) ($F_5$ in Figure~\ref{fig:prototype:features}).
For implementation, we used the COCA dataset, where we precomputed the number of occurrences of n-grams (1$\leq$n$\leq$4) from the dataset, recording the total count and count per year. This data was stored in our database and retrieved when users submitted queries.

\subsection{\system{} Implementation}
We implemented \system{} using JavaScript, HTML, and CSS, building upon the CoAuthor interface~\citep{coauthor}. 
We used a Flask server for the backend to preprocess requests from the front-end and forward these requests to the necessary destinations.
All log data was stored on a local server and pseudonymized using identifiers recognizable only by the authors.
\section{User Study}
\label{section:main}

\rev{
We conducted a user study with 22 NNESs using \system{} to gain insights into how information aids can be effectively integrated into AI paraphrasing tools. 
Concretely, we aimed to address the following research questions:

\begin{enumerate}
    \item [\textbf{RQ1:}] How do NNESs interact with information aids to assess and select AI paraphrase suggestions in \system{}?
    \item [\textbf{RQ2:}] What impact does \system{} have on NNESs’ user experience and writing performance?
    \item [\textbf{RQ3:}] What kinds of interactions and interface features do NNESs wish to have in \system{}?
\end{enumerate}

% Rather than conducting a comparative study with baselines to evaluate the effectiveness of the system, we opted for a single in-depth user study to closely observe NNESs’ interactions with \system{}, aiming to develop a holistic understanding of how information aids are used in paraphrasing context~\citep{10.1145/3581641.3584054, 10.1145/3544548.3581388, 10.1145/3643834.3660705}.
}

\begingroup
\setlength{\tabcolsep}{8pt} % Default value: 6pt
\begin{table}[t]
\centering
\scalebox{0.8}{ % adjust the scale factor as needed
\begin{tabular}{c c l} %{@{}lcllll@{}}
\toprule
 \begin{tabular}[c]{@{}c@{}}Participant \\ ID\end{tabular}  & \begin{tabular}[c]{@{}c@{}}English \\ Proficiency\end{tabular} & Academic Status \\ 
 \midrule
P1 & A2 & Master's student \\
P2 & A2 & Undergraduate student \\
P3 & A2 & Undergraduate student \\
P4 & A2 & Undergraduate student\\
P5 & A2 & Undergraduate student\\
P6 & B1 & Undergraduate student\\
P7 & B1 & Master's student \\
P8 & B1 & Ph.D. student \\
P9 & B1 & Ph.D. student \\
P10 & B1 & Master's student \\ 
P11 & B1 & Master's student \\ 
P12 & B1 & Master's student \\
P13 & B1 & Undergraduate student\\ 
P14 & B1 & Ph.D. student \\
P15 & B1 & Undergraduate student\\ 
P16 & B2 & Master's student \\ 
P17 & B2 & Master's student \\ 
P18 & B2 & Ph.D. student \\ 
P19 & B2 & Undergraduate student\\ 
P20 & B2 & Undergraduate student\\ 
P21 & B2 & Master's student \\ 
P22 & B2 & Undergraduate student\\ 
\bottomrule
\end{tabular}
}
\vspace{1em}
\caption{Detailed background information of the interview participants in the main study (Section~\ref{section:main}). The first language of all participants was Korean. For English proficiency, we use self-reported CEFR levels: A1 (beginner) < A2 (elementary) < B1 (intermediate) < B2 (upper intermediate) < C1 (advanced) < C2 (proficient).}
\Description{The table shows the background information of the interview participants in the main study. The first row is participant ID, second is English Proficiency ranging from A2 to B2, and the last is the academic status of participants.}
\vspace{-2em}
\label{tab:main:participants}
\vspace{-.5em}
\end{table}
\endgroup

\subsection{Participants}
\label{section:main:participants}
We recruited 22 NNESs through advertisement posts in university online communities. 
We targeted university students as they are among the most representative NNESs who often write in English, especially emails in academic settings.
We had two inclusion criteria: (i)~one’s self-assessed English proficiency level is below or equal to B2 (upper intermediate) according to CEFR~\cite{council2001common} measurement,\footnote{CEFR levels are defined as basic (A1: Beginner; A2: Elementary), independent (B1: Intermediate; B2: Upper intermediate), and proficient (C1: Advanced; C2: Proficient).} and (ii)~one is familiar with at least one AI paraphrasing tool. 
This selection was made to explore the behaviors and needs of users with limited English proficiency and to prevent the novelty effect often associated with first-time use of paraphrasing tools. 
Additionally, we asked applicants to submit their email excerpts optionally, which were utilized to create email scenarios used in the user study.

\rev{After applying these criteria, we randomly selected 22 participants~(Age=19-34, Mean=24.8, Std=3.9; Female=12, Male=10).}\footnote{\rev{Among the 25 initially recruited participants, we conducted a pilot study with the first three to address potential technical issues and improve the tool tutorial.}} 
All participants’ first language was Korean. Ten were undergraduates, eight were master's, and four were PhD students. Five had an English level of A2 (elementary), ten had B1 (intermediate), and seven had B2 (upper intermediate). The detailed demographic data of the participants are shown in Table~\ref{tab:main:participants}. 
The study lasted a maximum of 117 minutes (M=105.55, SD=10.9), and participants were compensated with approximately 27 USD (40,000 KRW). 

\subsection{Study Procedure}
\begin{figure*}[t!]
    \centering
    \includegraphics[width=\linewidth]{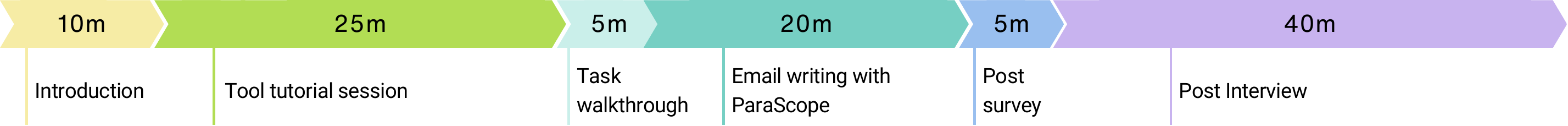}
    \caption{Overview of the user study procedure. After familiarizing themselves with each information aid in the tutorial session, participants wrote an open-ended academic email with \system{}. After the writing, participants engaged in a post-survey and semi-structured interview about their experiences.
    }
    \Description{
    This figure illustrates the overview of the user study procedure. It starts with a 10-minute introduction and tool tutorial, followed by the 25-minute tool tutorial. In the writing task, users went through a 20-minute email writing session with the given tool. After the task, participants complete the 5-minute post-survey, which is followed by a 40-minute interview session.
    }
    \label{fig:main:procedure}
    \vspace{-.5em}
\end{figure*}

The entire process of the user study is shown in Figure~\ref{fig:main:procedure}. 
Every participant took part in the study online via Zoom. 
The study began with a brief introduction to the research and the signing of informed consent forms. 
After the introduction, participants completed a 25-minute-long tutorial task. 
During the tutorial, one of the authors introduced the tool and each information aid's functionality, and then participants completed a series of subtasks selecting the best-paraphrased suggestion in the context of email writing, given a pre-written sentence and four paraphrased suggestions with \system{}.\footnote{Participants were instructed to access the link to the tool with an incognito window blocking any third-party writing tools. This was to accurately log the information aids participants use and their interactions with the paraphrased suggestions and information aids.} To ensure participants familiarized themselves with each information aid, only one of the five information aids was available for each subtask, and they could ask questions about the tools during the tutorial. 

After the tutorial, participants proceeded to the email writing task, where they were given an open-ended email writing task with one of the academic scenarios we prepared (Appendix~\ref{appendix:main:emailscenarios} lists the scenarios). These scenarios were designed to be realistic so that college students could easily relate to them (e.g., inquiring about grade corrections or requesting to audit courses). The scenarios were selected from the email excerpts from the recruitment survey. 
The scenarios assignment was counterbalanced across participants to minimize bias.
The goal of the user study was to investigate how participants used the paraphrasing functionality and the information aids in an open-ended email writing task. 
As such, we did not require users to engage with every feature; participants were informed that they could freely choose which information aid to use.
During the task, the researcher observed and noted their paraphrasing activities and asked about them during the interview.

After the tasks, participants completed a post-survey. The survey included questions about their overall experience using each information aid feature in \system{} and users' overall ratings of the quality and diversity of the paraphrased suggestions. 
Finally, we conducted semi-structured interviews with all participants on Zoom. 
During the interview, we asked participants to explain their paraphrasing process using \system{} and share their experiences and suggestions with the paraphrasing functionality and the information aids. 
Under the consent of the participants, all interviews were recorded and later transcribed for further analysis.

\subsection{Analysis}

\subsubsection{Quantitative Analysis}
To investigate participants' usage patterns with \system{}, we conducted a quantitative analysis using the interaction logs collected from the email writing task. 
We counted the number of paraphrasing events per each writing session, as well as the number of associated information aid usages for each paraphrasing event. 
We analyzed the acceptance rate of paraphrased suggestions using the Chi-Square test.
To identify differences among participants with varying levels of English proficiency, we analyzed the average number of paraphrasing events and information aids used using the Kruskal-Wallis test for overall comparisons and the Mann-Whitney U test for pairwise comparisons.
\rev{These non-parametric tests were selected as the Shapiro-Wilk test revealed the data was non-parametric.}
We also investigate the usage trends of each information aid over time using the timestamped information aid usage logs per each paraphrasing event.

Additionally, we evaluated paraphrased sentence pairs (original sentence vs. selected paraphrase) collected in the user study through both automatic human evaluations to investigate whether using information aids has to do with sentence quality improvement. 
For the automatic evaluation, we measured grammaticality using the LanguageTool API~\cite{languagetoolapi} following a previous work~\cite{coauthor} to compute the number of spelling and grammar errors per sentence. The scores were averaged across sentences to compare the quality of sentences before and after paraphrasing.
For human evaluation, we focused on two key aspects: fluency and politeness for contextual appropriateness. 
We presented human evaluators with sentence pairs (original sentence vs. selected paraphrase) and asked them to choose which sentence ``sounds more natural'' (fluency) and ``is more polite'' (politeness).
We included the question about politeness because the email tasks in the user study involved making requests to a person in authority, and each task description explicitly instructed participants to write the email \textit{politely}.
We recruited 24 native English speakers from Prolific~\cite{prolific} and compensated \pounds 0.75, which corresponds to \pounds 9 per hour. 
Each evaluator evaluated 28 sentence pairs (14 for fluency and 14 for politeness), and three different evaluators assessed each sentence pair. 
We analyzed the pairwise judgments using the Wilcoxon signed-rank test. 
\rev{We analyzed the pairwise judgments using the Wilcoxon signed-rank test, a non-parametric alternative to the paired t-test, as confirmed by the Shapiro-Wilk test.}

\subsubsection{Qualitative Analysis}
\rev{
We transcribed the semi-structured interviews and analyzed them using the constant comparative method~\citep{strauss2017discovery}. 
Two authors independently open-coded two interview samples to identify key concepts and patterns.
Axial coding was then performed to link these patterns~\citep{corbin2014basics}, resulting in an initial codebook.
The first author coded the remaining interview data while continuously refining the codebook.
Upon completing the first coding round, another author coded one interview sample based on the updated codebook for verification, and any issues were resolved through discussion. 
Throughout the process, all authors regularly discussed emerging themes while triangulating the interview data with quantitative analyses. 
We include the full codebook in Appendix~\ref{appendix:study:codebook}.
}

All interviews and qualitative analysis of interview transcriptions were conducted in Korean, and the selected quotes were translated into English. The authors reviewed all translations to ensure accuracy and preserve the integrity of participants' original statements.

\section{Results}
\label{section:findings}

\rev{
We present our findings from the user study by explaining observations of NNESs' interaction patterns with information aids (RQ1), potential impacts of integrating information aids on NNESs' user experience and writing performance (RQ2), and participants' suggestions on the system UI and interaction method (RQ3). We summarize our main findings for three research questions below:
}

\rev{
\begin{itemize}
    \item [\textbf{RQ1:}] Participants favored simple global features, especially among lower-proficiency users. While local features were used less often, they helped resolve difficult choices.
    \item [\textbf{RQ2:}] Using information aids improved perceived confidence and efficiency, but sometimes caused information overload. Participants also reported potential for increased autonomy in the process and language learning.
    \item [\textbf{RQ3:}] Participants suggested greater transparency in information aids, interface personalization, editable suggestions, and inclusion of the original sentence for comparison.
\end{itemize}

While it is not the primary focus of our study, we note that users expressed general satisfaction with the quality~(Mean=4.23, Std=0.43) and diversity (Mean=4.36, Std=0.66) of suggestions from our tool (in a 5-point Likert scale, 1=Strongly Disagree, 5=Strongly Agree). This satisfaction with paraphrased suggestions indicates that participants generally did not experience usability problems in their writing task in terms of the paraphrasing outcomes. 
}

% \subsection{User Motivations for Seeking Writing Support}
% We first uncover NNESs' motivations for paraphrasing and using information aids identified from the post-interview.

% %
% \subsubsection{\textbf{Motivations for paraphrasing}}
% Post-interview analyses revealed that NNESs used \system{} for two purposes. First, eight participants mentioned they relied on the paraphrasing capability of \system{} to compose sentences, anticipating AI would generate a polished, ready version of their sentences. P20 noted: \textit{``The paraphrasing tool helped compose sentences. I initially wrote some keywords without worrying about grammatical correctness or fluency, relying on the tool to turn my inputs into coherent and fluent sentences.’’} 
% Second, ten participants mentioned using paraphrasing tools to improve the quality of sentences they have written. P8 asked for paraphrases to see if her sentences were \textit{``okay enough,’’} where she \textit{``compared my sentences with paraphrased suggestions.’’} Upon finding a better paraphrase, she replaced her original sentence with the suggestion.
% These motivations for using paraphrasing tools align with findings from prior work that studied NNESs' paraphrasing behaviors~\citep{rewriting}.

\subsection{NNESs' Interaction with Information Aids}
\label{section:findings:approach}

\rev{
We explore how participants interacted with information aids during paraphrasing (RQ1). 
We describe relevant quantitative findings with qualitative insights.

Overall, we collected a total of 258 paraphrasing events from the user study. The average paraphrasing events made per user was 11.73 (Std=5.55, Max=27, Min=4, Median=12.0), with an acceptance rate (Accept = users accept one of the AI paraphrased suggestions, Reject = close without selecting) of 61.63\%. 
Participants used at least one information aid in 193 (=74.81\% of all events) paraphrasing events. 
The average number of information aids used by participants was 1.93 (Std=0.93, Median=2.0). 
}

\subsubsection{\textbf{Motivations for using information aids}}
\label{section:findings:reasons:2}
During the post-interview, participants noted two primary motivations for using information aids: (i)~they were uncertain about the suggestion quality and used the features to \emph{assess} and \emph{rank} the suggestions; 
(ii)~participants with a specific decision in mind referred to the features to \emph{validate} their choice. 
When leveraging the features, participants considered three aspects of suggestions in their decision-making process: (i)~whether suggestions sound overall natural in English, (ii)~ the appropriateness of the tone, and (iii)~real-world applicability of words/expressions. This observation aligns with our formative study findings (\S\ref{section:system:feature}), showing that these aspects are commonly considered in NNESs’ writing activities.

\subsubsection{\textbf{\rev{Preference for global features}}}
\begin{figure}[t!]
    \centering
    \includegraphics[width=0.9\linewidth]{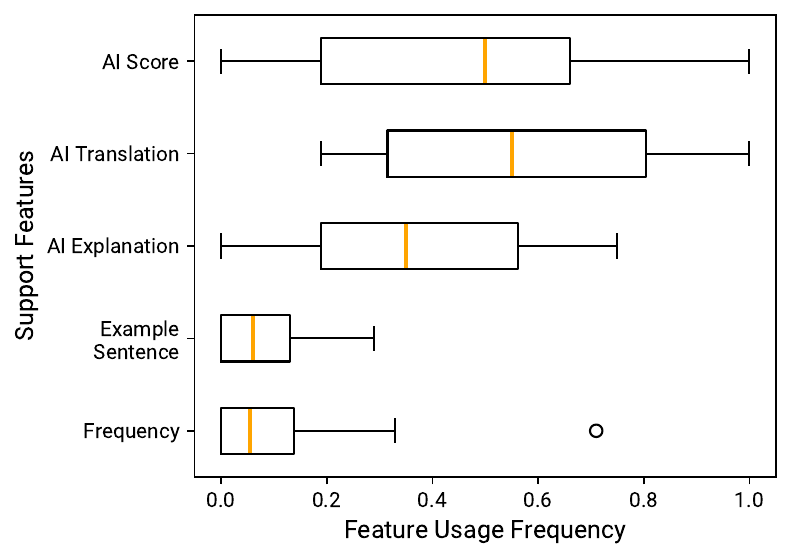}
    \vspace{-1em}
    \caption{Feature usage frequencies for each information aid. The orange bar indicates the median value. While AI Score (M=0.46, SD=0.33), AI Translation (M=0.56, SD=0.27), and AI Explanation (M=0.37, SD=0.23) were frequently used in a paraphrasing event, usage frequencies of Example Sentence (M=0.08, SD=0.1) and Frequency (M=0.11, SD=0.17) were comparatively low.}
    \vspace{-1em}
    \label{fig:findings:freq}
    \Description{
    This figure shows the box plots of feature usage frequencies for each support feature. AI Score, AI Translation, and AI Explanation are frequently used, while Example Sentence and Frequency have relatively low frequency values.
    }
\end{figure}

% More frequent usage of global features than local features
\label{section:findings:approach:1}
Across all events, we observed significantly higher usage frequencies of \bglobal{} features (\fscore{}, \ftranslation{}, \fexplanation{}) 
compared to \blocal{} features (\fexample{}, \ffrequency{}) ($\chi^2=179.89, p<.001$).
Among these, \ftranslation{} was the most frequently used feature (63.18\% of all events), 
followed by \fscore{} (47.67\%) and \fexplanation{} (40.70\%). 
In contrast, \ffrequency{} (8.91\%) and \fexample{} (6.98\%) were rarely used. 
To account for potential biases from participants with higher paraphrasing event counts, we further analyzed feature usage frequencies averaged across individual participants. 
Figure~\ref{fig:findings:freq} illustrates these results, confirming a significant difference among feature groups ($\chi^2=47.68, p<.001$). 
Pairwise comparisons using the Mann-Whitney U test showed that all global features were used significantly more frequently than local features ($p<.001$ for all comparisons). 

Post-interview analysis revealed that \rev{\textbf{participants’ preferences were largely driven by the simplicity of the information and the ease of interaction.}}
Notably, 15 out of 22 participants mentioned using information aids to quickly and roughly assess suggestion quality, allowing them to narrow down multiple suggestions to a manageable set for focused evaluation. 
For example, P3 described using features to \textit{“initially prioritize my preferences”}, while P12 noted a similar approach of \textit{“screening out”} undesirable suggestions to focus on the most viable options.

Eight participants \rev{\textbf{particularly favored \fscore{} and \ftranslation{} due to their straightforward presentation}}, which is effective in quickly assessing suggestion qualities. 
P10 said, \textit{``AI Score is presented as a number, and the AI Translation is in my native language, so they're easy to understand and efficient.’’}
Conversely, \rev{\textbf{local features were less preferred because they were perceived as time-consuming and effortful to use}}. For both \fexample{} and \ffrequency{}, eight participants reported that deciding on search queries and manually typing them added extra effort. P1 mentioned: \textit{``To search, I have to decide what to look for and type it in---it takes time and effort, so I didn’t use it much.’’}
\fexample{} was further criticized for providing indirect and harder-to-comprehend information, as noted by six participants. 
P7 noted, \textit{``It doesn’t give me a clear answer. I have to read, think, and judge the information, which makes it difficult to use.’’}

% Participants also suggested that the task context influenced their feature usage. The study required participants to complete an academic email-writing task within a constrained timeframe, which may have discouraged the exploration of more time-intensive features. P6 mentioned: \textit{``I’d use it if I had more time, if the email was really important to me, or if I were trying to learn English. But since this was a user study and I needed to finish the task, I focused on completing it rather than learning.''}

Lastly, participants \rev{\textbf{had mixed opinions on \fexplanation{} in terms of its simplicity}}. Eight participants appreciated its explicit and actionable information, as P2 noted: \textit{``For example sentences, I need to read and interpret the information myself. But AI Explanation directly tells me, ‘This is often used here,’ or ‘This might work better,’ which makes it easier because I don’t have to think as much.''}
However, four participants found \fexplanation{} cumbersome due to its length and text-heavy format: \textit{``I needed to make quick decisions on suggestions, but since it’s presented in long paragraphs, it wasn’t easy to skim. If it were in bullet points, it would have been faster and more convenient.’’} (P8).

\subsubsection{\textbf{More reliance on information aids for lower proficiency users}}
\label{section:findings:approach:3}
\begin{figure}[t!]
    \centering
    \includegraphics[width=0.95\linewidth]{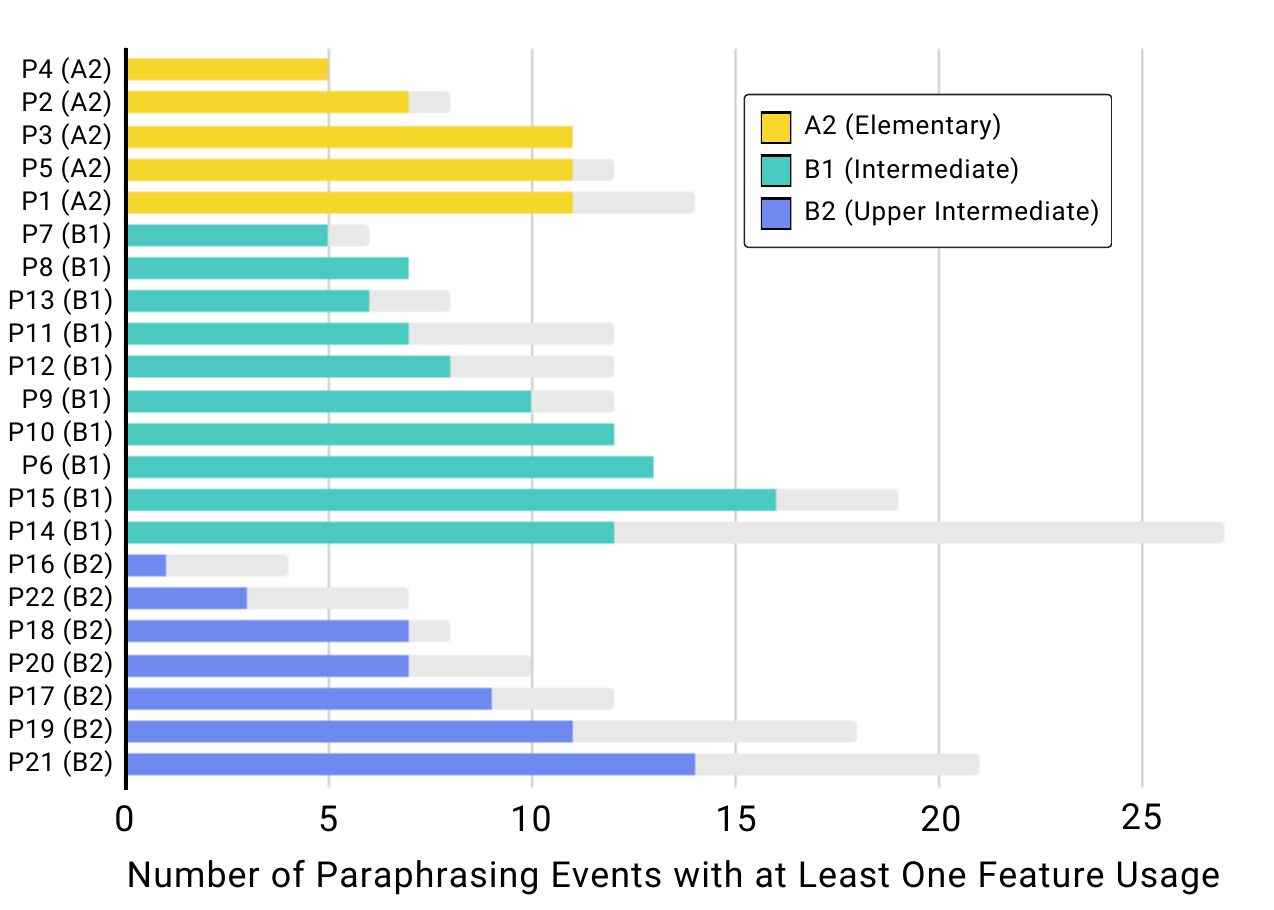}
    % \vspace{-1.5em}
    \caption{Number of paraphrasing events per user. Out of the total paraphrasing events (gray bars), we show the number of events with at least one support feature usage (colored bars).}
    \label{fig:findings:peruser}
    \Description{
    This figure shows the number of paraphrasing events per participant in a stacked bar plot. The total paraphrasing events by each user is colored gray, and the number of events with at least one support feature usage is colored. For the elementary level users, it is yellow, intermediate users green, and upper intermediate users blue.
    }
\end{figure}

\begin{figure}[t!]
    \centering
    \includegraphics[width=0.9\linewidth]{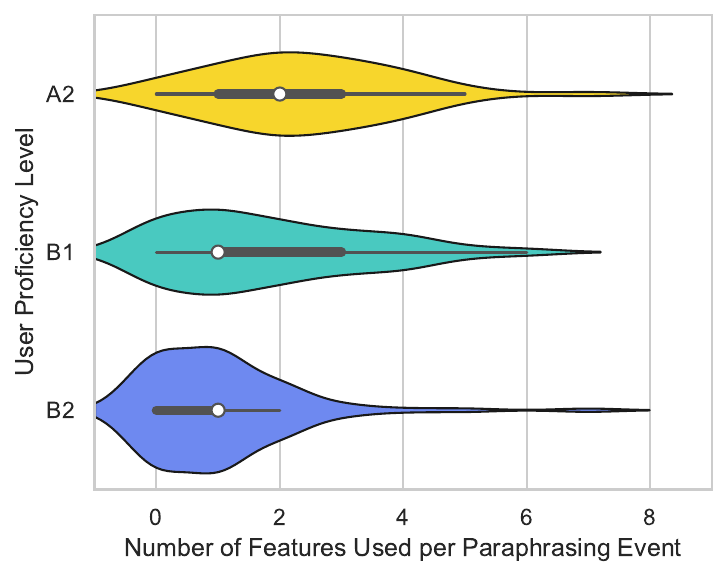}
    \vspace{-.4cm}
    \caption{Distribution of the number of features used in one paraphrasing event by user proficiency levels: A2 (elementary), B1 (intermediate), and B2 (upper intermediate). The white dot represents the median and the thick gray bar represents the interquartile range. Values outside the thin line are considered outliers. (A2: M=2.4, SD=0.61; B1: M=1.96, SD=0.78; B2: M=1.17, SD=0.35)}
    \vspace{-.2cm}
    \label{fig:findings:violin}
    \Description{
    This figure shows the distribution of the number of features used in a single paraphrasing event in a violin plot. We show three violin plots per user proficiency level. The violin plot for the elementary user is the most flat, with an average number of feature usage of 2.4; the plot for the intermediate user has an average value of 1.96; the plot for the upper intermediate user is the widest, with its average value of 1.17.
    }
\end{figure}

We discovered that, in general, participants with lower English proficiency tend to rely more heavily on information aids when selecting AI paraphrases compared to those with higher English proficiency. 
Figure~\ref{fig:findings:peruser} shows the number of paraphrasing events per participant, with at least one feature usage. Out of 258 instances of paraphrasing, at least one information aid was used in 193 cases (=74.81\%). Notably, the proportion of paraphrasing events with information aids was significantly higher among participants with lower proficiency levels ($\chi^2_{(2)}=9.60, p=0.0082$; A2: M=91.55, SD=9.05; B1: M=80.69\%, SD=18.96; B2: M=47.57\%, SD=32.51). 
We also statistically analyzed the number of information aids used in each paraphrasing event by user proficiency levels. The average number of features used was higher as user proficiency got lower: A2 users, on average, used 2.4 features (SD=0.61), B1 used 1.96 features (SD=0.78), and B2 used 1.17 features (SD=0.35), as shown in Figure~\ref{fig:findings:violin}. Kruskal-Wallis test also revealed that this trend is statistically significant ($\chi^2_{(2)}=27.79, p<.001$). 

From the pairwise statistical analysis of the feature usage frequency data using Mann Whitney U test, we noticed the English proficiency of a participant correlated with AI Score usage frequency: \rev{\textbf{participants with lower English proficiency levels (A2, B1) were more likely to use \fscore{} more frequently than those with the highest proficiency levels (B2)}} (A2>B2: $U=2.5, p<.05$; B1>B2: $U=12.0, p<.05$; A2: $M=0.72, SD=0.32$; B1: $M=0.51, SD=0.29$, B2: $M=0.19, SD=0.21$). 
Similarly, during the post-interview, we discovered instances where \rev{\textbf{participants with lower English proficiency put more trust in \fscore{}}}. P1, whose English proficiency is A2 (elementary), mentioned relying on \fscore{} rather than personal judgments, reasoning that \textit{``AI probably has a better understanding of English than I do.''} 
In contrast, P22, whose English proficiency is B2 (upper intermediate), was more critical of \fscore{} and preferred making their own evaluations: \textit{``It was hard to understand the criteria the AI used to generate the score. I thought it would be better to rely on my own judgment instead.''}

\subsubsection{\rev{\textbf{Frequent usage but lower acceptance rate for \ftranslation{}}}}
\label{section:findings:approach:2}
Paraphrasing events with information aids had a higher acceptance rate of 67.36\% than those without information aids (44.62\%) ($\chi^2=14.58, p=0.0019$). 
Interestingly, \textbf{\rev{while \ftranslation{} was the most frequently used feature}} (usage frequency of 63.18\% in \S\ref{section:findings:approach:1}), \textbf{\rev{it was not associated with higher acceptance rates}} of suggestions ($\chi^2=0.79, p=0.375$).
In contrast, \textbf{\rev{\fscore{} and \fexplanation{} showed significant correlation with higher acceptance rates}} of suggestions ($\chi^2=6.34, p=0.012$ for \fscore{}, $\chi^2=10.01, p=0.0016$ for \fexplanation{}).

These results suggest that \ftranslation{}'s utility might lie more in comprehending suggestions rather than assisting in comparing suggestions, i.e., deciding which one is the best. While it may help understand the semantic meaning of suggestions as indicated by prior works~\citep{Kobayashi1992EffectsOF, chamot1987study}, it may not be as critical for determining fluency of suggestions or appropriateness in tone, which were major considerations in NNESs' decision-making process as identified in \S\ref{section:findings:reasons:2}. 
Conversely, as \fscore{} enables direct quantitative comparison among suggestions, it likely supports users in objectively evaluating the relative quality of suggestions. Similarly, \fexplanation{}, by explicitly explaining tone appropriateness, may guide users in identifying subtle tonal differences and aligning suggestions with the intended communicative goals. 

% \begin{figure*}[t!]
%     \centering
%     \begin{minipage}[b]{0.45\linewidth}
%         \centering
%         \includegraphics[width=0.8\linewidth]{figs/resources/feature_usage_per_user.pdf}
%         \vspace{-1.5em}
%         \caption{Number of paraphrasing events per user. Out of the total paraphrasing events (gray bars), we show the number of events with at least one support feature usage (colored bars).}
%         \label{fig:findings:peruser}
%     \end{minipage}
%     \hfill
%     \begin{minipage}[b]{0.45\linewidth}
%         \centering
%         \includegraphics[width=0.8\linewidth]{figs/resources/feature_usage_violin.pdf}
%         \vspace{-.4cm}
%         \caption{Distribution of the number of features used in one paraphrasing event by user proficiency levels: A2 (elementary), B1 (intermediate), and B2 (upper intermediate). The white dot represents the median and the thick gray bar represents the interquartile range. Values outside the thin line are considered outliers. (A2: M=2.4, SD=0.61; B1: M=1.96, SD=0.78; B2: M=1.17, SD=0.35)}
%         \vspace{-.2cm}
%         \label{fig:findings:violin}
%     \end{minipage}
% \end{figure*}


\begin{figure}[t!]
    \centering
    \includegraphics[width=\linewidth]{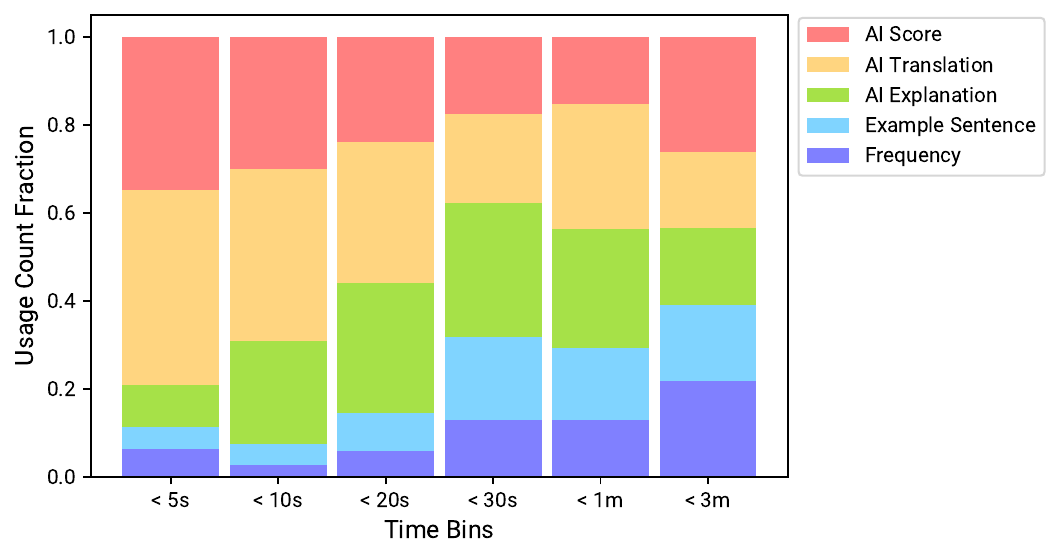}
    \caption{Each support feature's usage fraction in six phases (time bins) of a paraphrasing event. This figure visualizes which support features participants utilized at what time in a paraphrasing event. 
    Each feature use event is counted within each time bin across all paraphrasing events; for example, the AI Score event is counted to the `$<5s$' time bin if the AI Score was clicked two seconds after receiving suggestions.
    We determined the segmentation of time bins based on the distribution of feature-use events.}
    \label{fig:findings:timebin}
    \Description{
    This figure shows each support feature's usage fraction in six phases, in other words, time bins, of a paraphrasing event. For the first time bin, AI Score and AI Translation have large fraction values. From the second to the fifth bar, the portion of AI Score and AI Translation gradually decreases, while the fraction of AI Explanation, Example Sentence, and Frequency gradually increases. For the last time bin, Frequency and AI Score has the largest fraction.
    }
\end{figure}

\subsubsection{\textbf{Use of global features in early decision-making, with local features in later stages}}
\label{section:findings:approach:4}
While global features played a dominant role in the decision-making process, we observed an intriguing trend: local features became increasingly utilized in the later stages, as illustrated in Figure~\ref{fig:findings:timebin}. The figure provides a temporal breakdown of how different information aids were employed during a paraphrasing event. 
Notably, as users spent more time, the proportional usage of local features (\fexample{} and \ffrequency{}) tended to rise.

Insights from the post-interview revealed that participants \rev{\textbf{relied on local features to further investigate suggestions or as tie-breakers when initial aids were insufficient.}}
For example, five participants reported using \fexample{} when the explanation provided by \fexplanation{} left them uncertain about a suggestion’s appropriateness.
P15 explained: \textit{``Even if the AI explanation explicitly states the situations where a suggestion can be used, it doesn’t always cover every possible context. When the explanation seemed ambiguous, I checked the example sentences to confirm the appropriate usage of the expression.’’}
Similarly, four participants mentioned turning to Frequency as a tie-breaker. P6 described their process:
\textit{``After looking at the score and the AI explanation, if I still couldn’t decide on the best option, I used Frequency to choose the expression that was more commonly used.’’}
In addition, the \fscore{} feature was also used for tie-breaking, as mentioned by six participants. For instance, P5 noted: \textit{``If I was still unsure which of two sentences to use, I thought, perhaps the one with the higher numerical value would be better.’’}

\rev{
\subsubsection{\textbf{Summary of findings.}}
NNESs used information aids to assess fluency, tone appropriateness, and the real-world applicability of expressions. They preferred global features (\fscore{}, \ftranslation{}, and \fexplanation{}) for their simplicity and efficiency, especially among lower-proficiency users.  
Despite being the most frequently used, \ftranslation{} did not correlate with higher acceptance rates, whereas \fscore{} and \fexplanation{} did.
Local features were less frequently used overall but played a role as tie-breakers in the later stages of suggestion evaluation.
}

\vspace{1em}
\subsection{Impact on User Experience and Writing Performance}
\label{section:findings:impact}
% While the main purpose of integrating information aids within an interface was to analyze user engagements with different information aids on an equal footing, we observed that users were, in general, satisfied with the features being integrated into a single interface.
\rev{
We describe perceived user experiences of using \system{} in the email writing task among NNESs (RQ2). 
We report findings from qualitatively analyzing post-interviews, as well as quantitative findings from human evaluation results of user-written sentences.
}

\subsubsection{\textbf{Informed decision-making with multiple information aids}}
\label{section:findings:impact:2}
Having multiple information aids appeared to \textbf{\rev{benefit participants by enhancing confidence in the decision-making process}}, as mentioned by 13 participants.
For example, P2 remarked: \textit{``Having access to various types of information that either confirmed or refuted the suitability of a suggestion made me feel more reassured and confident in my choice.''} 
Eight participants \textbf{\rev{described their decision-making as a multi-step verification process, using diverse information aids to offer a richer perspective.}} 
For instance, P1 described their approach as starting with the AI Score for an initial evaluation and then refining their decision by cross-referencing other features: \textit{``I used the AI Score to gain initial confidence in the suggestion, but still felt uncertain. So, I checked other features, which gradually strengthened my confidence.''}
P22 also mentioned using multiple features leads to more accurate judgments: \textit{``Relying on just one feature might lead to a hasty decision, but using multiple features to make a comprehensive judgment improves the accuracy of my choices.''}
Additionally, \textbf{\rev{seven participants mentioned using one feature to better understand or validate another.}} 
For instance, P11 elaborated on how combining feature clarified the \fscore{}'s scoring rationale: 
\textit{``When I only looked at the AI Score, I doubted its reliability and couldn’t understand why it rated a suggestion the highest. But when I checked the AI Translation and Explanation, it made sense, and I could accept the reasoning behind the score.''}

\subsubsection{\textbf{\rev{Efficient information-finding process, but possible information overload}}}
\label{section:findings:impact:1}
Ten out of 22 participants mentioned that having all features in the same interface as the paraphrasing tool \textbf{\rev{enabled easy and fast access to necessary information without switching between multiple windows or tools.}} P21 stated, \textit{``Typically, my writing process involves constantly switching between windows: composing, searching dictionaries, and consulting ChatGPT. This system consolidates these actions within one platform, significantly aiding my workflow.''} 

\rev{
Participants expressed mixed opinions about the cognitive impact of this integration.
Twelve participants noted that \textbf{having multiple features in one place allowed them to delegate part of the decision-making process} to the system. 
For example, P15 remarked, \textit{``Rather than deliberating over each suggestion on my own, I find it more convenient to quickly scan the text and let the features handle the finer details of decision-making.''} 
Similarly, P1 described how their efficiency improved over time: \textit{``As I became more accustomed to using the features, my ability to analyze and make decisions became faster.''} 
In contrast, five participants 
found participants found that the \textbf{abundance of information aids led to stress} during decision-making, as they felt pressured to process and use every available resource. For instance, P8 noted: \textit{``There was so much to refer to that it actually got in the way of making a decision.''} 
Given all the information aids, P6 felt responsible for checking every suggestion and feature presented in the system: \textit{``I feel compelled to use every information aid, even for sentences I could easily move on from. I don't think this process is efficient, but I am nonetheless convinced that it improves writing.''}
}

\subsubsection{\textbf{Enhanced autonomy in decision-making process}}
\label{section:findings:impact:3}
Interestingly, two participants reported that \textbf{\rev{utilizing information aids increased their sense of autonomy in decision-making.}} P1 mentioned, \textit{``When there was only AI Score, I relied on it, thinking it would make better choices than me. However, engaging with all five features changed my perspective; I felt more in control of my decisions, as the features seemed to support rather than dictate my choices.''} P18 highlighted that this process also augmented their sense of ownership over the final text: \textit{``Utilizing features to understand how suggestions might sound and integrating these suggestions into my text, made me feel more actively involved in the writing process. It felt as if I were crafting the text myself.''}

\subsubsection{\textbf{Potential for language learning}}
\label{section:findings:impact:4}
Fourteen participants anticipated a potential learning impact with the integrated features. 
This expectation appeared to stem from the belief that \rev{\textbf{increased interaction with the features could lead to learning experiences.}} 
P15 highlighted that \textit{``engaging in the comparative analysis of suggestions’ various aspects could deepen understanding in English and promote thorough learning.''} 
Similarly, P13 noted that \textit{``the diverse insights provided by each feature enriched my engagement with English.''} Moreover, P22, majoring in English education, noted that the tool could particularly benefit language learners with low proficiency, as \textit{``the tool provides explanations for suggestions, enhancing accessibility to information needed for learning English.''}
P1 nominated \fexample{} for helpful in learning English: \textit{``Example Sentence, unlike other features that provide direct hints, allows for the indirect exploration of various sentences and fosters thoughtful consideration.''} 

% Participants also suggested that the task context influenced their feature usage. The study required participants to complete an academic email-writing task within a constrained timeframe, which may have discouraged the exploration of more time-intensive features. P6 mentioned: \textit{``I’d use it if I had more time, if the email was really important to me, or if I were trying to learn English. But since this was a user study and I needed to finish the task, I focused on completing it rather than learning.''}

\subsubsection{\textbf{Sentence Quality Improvements.}}
\label{section:findings:eval:quality}

\begin{figure}[t!]
    \centering
    \includegraphics[width=.85\linewidth]{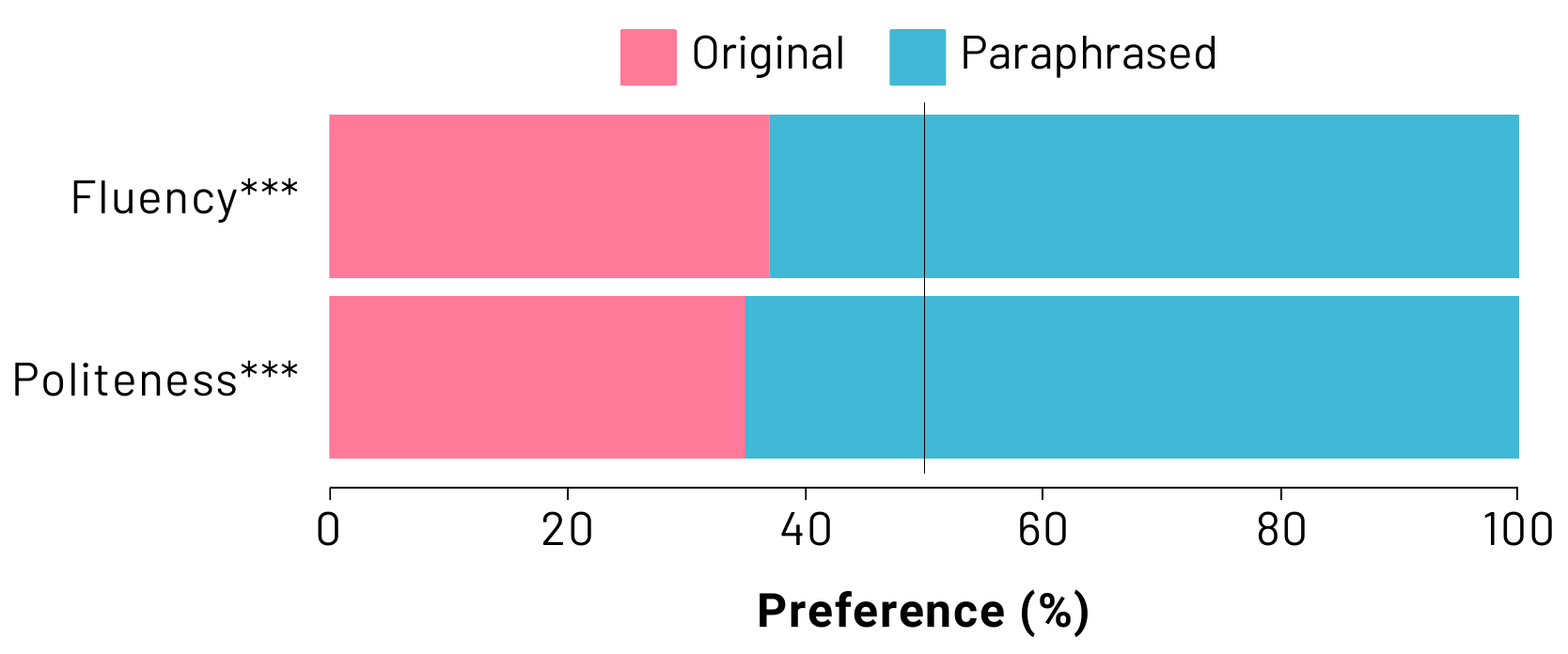}
    \caption{
    Human evaluation results of 112 paraphrased sentence pairs (original vs. selected paraphrase) composed in the user study. Among 336 comparisons, 65\% of evaluators rated the paraphrases as more fluent than the original sentences, and 67\% rated them as more polite. Overall, the paraphrased sentences showed a statistically significant improvement (***$p$ < .001).}
    \label{fig:findings:humaneval}
    \Description{This figure shows a bar chart comparing human preferences for fluency and politeness between original and paraphrased sentences. Paraphrased sentences were rated higher in both categories.}
\end{figure}

To examine whether paraphrasing with \system{} improved writing quality, we evaluated 112 sentence pairs (original vs. paraphrased) using both human judgments and automated metrics. 
Among 112 pairs, 100 pairs were paraphrased using information aids, while 12 were paraphrased without using information aids.
As a result, we observed an \textbf{overall improvement in sentence quality after paraphrasing}. 
Figure~\ref{fig:findings:humaneval} presents the human evaluation results of the pairwise comparison between NNESs' original and paraphrased sentences. Most evaluators preferred paraphrased sentences regarding fluency and politeness ($p$<.001). Similarly, in the automatic evaluation, the number of grammatical errors in the original sentences (M=0.42, SD=0.66) decreased significantly after paraphrasing (M=0.13, SD=0.37). 
These findings suggest that \system{}, which integrates information aids into the paraphrasing process, can help users select higher-quality suggestions. However, the small number of sentences paraphrased without aids (N=12) limits direct comparisons and warrants further investigation.
% While the primary focus of our study is to compare NNESs' interaction paraphrasing strategies, this finding suggests that AI paraphrasing tools can enhance NNESs' writing.

% In addition, we computed the grammatical errors of pairs grouped by information aids usage (used N=100; unused N=12). Sentences paraphrased without features initially had fewer errors (M=0.25, SD=0.60) than feature-assisted ones (M=0.44, SD=0.67). However, after paraphrasing, the sentences paraphrased using features had fewer errors (M=0.13, SD=0.36) than those paraphrased without features (M=0.18, SD=0.37). 
% While this observation may suggest that utilizing information aids could aid users in selecting better suggestions, the imbalance in the number of sentence pairs in the two conditions (N=100 vs. N=12) demands further research to validate these observations. 

\rev{
\subsubsection{\textbf{Summary of findings.}}
Participants reported that multiple information aids improved decision-making confidence, with many using them in combination for verification. The integrated interface supported efficient access but occasionally caused cognitive overload. Some users experienced increased autonomy and ownership over their writing, while others saw potential for language learning through deeper engagement with features. Human and automated evaluations showed improved fluency and politeness in paraphrased sentences, suggesting benefits for NNES' writing quality.
}

\subsection{\system{} Feature Requests}
\label{section:findings:suggestion}
We summarize key suggestions provided by participants for improving \system{}’s user experience (RQ3). We derive findings from qualitatively analyzing user interview data.

\subsubsection{\textbf{Explainability in numeric measures}}
\label{section:findings:suggestion:1}
Ten participants \textbf{\rev{emphasized the need for greater transparency and explainability in the \fscore{} feature.}} 
They wanted to understand \textit{``how the AI Score is calculated''} (P15) and \textit{``why some suggestions receive such high scores''} (P13).
Similarly, P11 expressed a preference for \fexplanation{} over \fscore{}, stating: \textit{``I didn’t understand the rationale behind the AI Score—it didn’t make sense on its own. Instead, I preferred reading the AI Explanation, as it helped me accurately assess whether suggestions aligned with my criteria.''} 
As such, P12 suggested \textit{``It would be helpful if the system explained the criteria used to calculate the scores and what those values represent.''}
A similar lack of transparency was noted for \ffrequency{}. P12 remarked: \textit{``It wasn’t clear what database the trends shown in Frequency were based on. For example, is this frequency higher because the word appears often in news articles? Knowing this would make the feature more helpful.''}

\subsubsection{\textbf{Personalization of information aids}}
\label{section:findings:suggestion:2}
Six participants suggested providing options to personalize the interface by selectively displaying features they used most often. 
This was related to the section (about information overload)
\rev{This suggestion was closely related to earlier concerns about information overload (see \S\ref{section:findings:impact:1}). Rather than displaying all information aids by default, participants \textbf{preferred a more streamlined interface tailored to their usage patterns.}}
For instance, P16 suggested compressing the interface to focus on frequently used features: \textit{``When using a tool repeatedly, people tend to stick to certain features. Instead of showing all features, it might be better to display only the ones I actively use.''} 
Participants also expressed interest in showing only frequently used features, with others available upon their needs. 
For instance, P15 suggested starting with \fscore{} as the default visible feature, with other aids revealed upon interaction: \textit{``Initially, only the AI Score could be shown, and if I’m curious, I could click the score to see related aids like AI Translation or AI Explanation.''}

\subsubsection{\textbf{Interactive Suggestion Refinement}}
\label{section:findings:suggestion:3}
Participants often wished to edit suggestions while the interface was open, highlighting two primary use cases. First, seven participants expressed a desire to \rev{\textbf{make minor adjustments to their original text based on the suggestions provided.}} For example, P17 explained: \textit{``I didn't want to completely rewrite my sentence; I aimed to keep the original structure intact while making minor adjustments to the words or expressions.''}
Second, five participants \rev{\textbf{preferred creating new sentences by integrating elements from multiple suggestions.}} For example, P9 shared an example of this process during the interview, where they started with the sentence `I figured out that there was a mistake' and received suggestions including `I discovered that an error had been made' and `I concluded that there was a miscalculation.' They integrated parts of these suggestions, revising their original sentence to `I discovered that there was a miscalculation.' 
In both cases, participants needed to type new sentences in the editor but found it challenging as the interface did not allow simultaneous editing and suggestion display. 

\subsubsection{\textbf{Incorporating Original Text for Comparison}}
\label{section:findings:suggestion:}
In our prototype, the original input text and its information aids were not displayed within the paraphrasing interface, following the design of existing tools~\cite{wordtune, langsmith}. 
However, 11 participants \rev{\textbf{suggested incorporating features like AI Score and AI Translation for the original input, as they frequently referenced their original sentence during decision-making.}} P6 explained: \textit{``It would be more convenient if my original input was displayed alongside the suggestions for direct comparison. Paraphrasing provides four candidates, but the original sentence is essentially a fifth option. Including translations or scores for the original would make it easier to evaluate all options equally.''} This feedback indicates that treating the original sentence as an integral part of the decision-making process could enhance the usability of paraphrasing tools.

\rev{
\subsubsection{\textbf{Summary of findings.}}
Participants suggested four key improvements to \system{}: (i) enhancing transparency in numeric features like AI Score and Frequency, (ii) enabling personalization by prioritizing frequently used aids, (iii) supporting flexible interaction with suggestions through in-place editing and combination, and (iv) displaying the original input with associated features for easier comparison. These suggestions highlight the need for more explainable, customizable, and interactive paraphrasing tools.
}
\section{Discussion and Design Implications}
\label{section:discussion}

With \system{}, we studied how NNESs assess AI-generated paraphrases with information aids. 
Based on our findings, we explore design implications for writing assistants to more effectively support NNESs and discuss broader impacts of the study.

\subsection{Design Implications}
In this section, we outline five design implications for writing systems tailored to NNESs, informed by our study findings.

\subsubsection{\textbf{Select information aids depending on user proficiency}}
We observed that participants had different usage patterns in information aids according to their English proficiency: participants with lower English proficiency used more information aids and relied significantly more on \fscore{} than those with higher proficiency (\S\ref{section:findings:approach:3}).
Based on the findings, we suggest that writing systems for NNESs should adapt the availability and presentation of information aids based on the user's proficiency level. For example, systems could provide multiple features, including \fscore{} by default for lower-proficiency users. Conversely, for higher-proficiency users, the interface could simplify the display by emphasizing fewer features only. 

In addition, writing tools could incorporate customization options, allowing users to set their own preferences for which information aids to display (\S\ref{section:findings:suggestion:1}). For instance, users could toggle specific features on or off based on their current needs or goals. 
Adapting to proficiency while allowing personalization of the tool interface could enhance usability and minimize cognitive overload by ensuring that only relevant features are presented.

\subsubsection{\textbf{Reveal information aids in stages}}
We observed from the study that while participants generally preferred global features, they later resorted to local features when global features alone were insufficient for making decisions (\S\ref{section:findings:approach:4}). 
Considering this usage pattern, progressive disclosure could be an effective design strategy for managing the presentation of information aids. Progressive disclosure sequentially reveals information and functionalities, presenting only essential features initially while keeping more complex or less frequently used options accessible but hidden~\cite{carroll1984training, nakatani1983soft, progressive}. This approach reduces cognitive load, prevents information overload, and encourages efficient interaction with the interface~\cite{10.1145/2858036.2858402, 10.1145/1518701.1519023, progressive}.
In the context of our system, progressive disclosure could involve displaying global features by default while allowing users to explore local features on demand. 
For example, activating secondary displays through interface controls (e.g., clicking buttons or toggling options) could provide users with additional details without cluttering the primary interface.

\subsubsection{\textbf{Connect between information aids}}
% explainability of information aids
While integrating different information aids in a single interface was helpful in a more informed decision-making process---for example, by checking AI Explanation to understand why AI Score is high, and checking Example Sentence to understand why AI Explanation says the sentence is polite---participants had to make extra effort to make connections between such features, such as deciding which term to search in Example Sentence (\S~\ref{section:findings:impact:2}). 
A more intuitive user interface design could effortlessly guide users through these information aids. 
For instance, AI Score might include an on-hover explanation or interactive icons detailing the reasoning behind certain scores. 
Moreover, visualizing attention scores~\cite{Bahdanau2014NeuralMT} of AI-driven features, such as AI Score or AI Explanation, could help users easily find key terms to delve deeper into. 
Attention scores measure the influence of different input tokens on a specific output token, and visualizing such influences (e.g., by highlighting each token with different opacity) could help users understand which input tokens significantly impact the model's output~\cite{alvarez-melis-jaakkola-2017-causal}. 
This support could be further enhanced by interaction designs that minimize user actions. 
For example, clicking the term with high relevance to the formality of a suggestion could automatically search the term in Example Sentence, streamlining the search process and reducing user effort.

\subsubsection{\textbf{Make AI suggestions and information aids interactive}}
In this study, \system{} provided static, non-editable suggestions and corresponding information aids, aligning with the design of existing tools~\citep{wordtune, langsmith, coauthor}. 
However, participants preferred editable suggestions and dynamic features that could adapt in response to their suggestion edits (\S\ref{section:findings:suggestion:2}).
This observation underlines the need to design paraphrasing tools that allow for a more flexible, interactive use of suggestions and information aids. 
We suggest that the suggestions provided within the interface be editable, so that users can modify suggestions within the interface. 
Previously static features like \fscore{}, \ftranslation{}, and \fexplanation{} could be updated in response to the user's editing suggestions. For example, a user may take the increase in \fscore{} as the user changes the word in its suggestion, as the word becomes more appropriate. Similarly, while we implemented \fexplanation{} in a way that it only compares the original and paraphrased sentences, the future \fexplanation{} could be implemented so that it tracks the user edits to the suggestion and explains the effects of edits accordingly. Providing immediate feedback on modifications with AI-powered information aids like this could transform the tool into a `what-if' analysis tool~\cite{whatiftool}, where users can experiment with sentence modifications and immediately observe the implications on quality. 

\subsubsection{\textbf{Provide information aids for both original and paraphrased texts}}
\label{section:discussion:original}
In our prototype, the original input text and its information aids were not displayed within the paraphrasing interface, following the design of existing tools~\cite{wordtune, langsmith}. 
However, participants frequently considered their original sentences during decision-making, often comparing them to paraphrased suggestions and rejecting suggestions when they felt the original was sufficient. 
Participants expressed the need for information aids, such as \fscore{} and \ftranslation{}, to be available for their original inputs to facilitate this comparison process (\S\ref{section:findings:suggestion:}).

This finding suggests a design implication that information aids should encompass both original and paraphrased texts. 
Implementation of information aids could be adjusted accordingly; for example, our current metric, ParaScore~\cite{parascore}, computes the AI Score of paraphrased text by measuring its semantic similarity and lexical divergence from the original, which is inapplicable to the original text itself. 
Instead, independent metrics, such as the metric for measuring fluency measurement~\cite{style20}, could offer a holistic view of both original and paraphrased texts.

\subsection{Information Aids in Broader NNES Writing Tasks}
Our study focused on designing information aids tailored to the specific needs of NNESs in email writing tasks (\S\ref{section:system:feature}). 
However, the design of such information aids will likely vary depending on the requirements of different writing contexts. Exploring task-specific information aids offers a promising direction for future research, as distinct writing tasks emphasize different aspects of language.

As identified in our study, email writing emphasizes tone, formality, and appropriateness~\citep{emailcommunication, varonis1985miscommunication}. 
In contrast, argumentative writing prioritizes clarity, conciseness, and logical coherence~\citep{coauthor}. 
Future research could investigate the most effective types of information aids for supporting these priorities. 
Such aids might identify vague or redundant sentences, suggest more precise vocabulary, or evaluate the logical structure of arguments. 
They could also highlight weak transitions between ideas or provide templates for organizing persuasive content.

Another important area for exploration is addressing the challenges of working with LLM-generated content.
Many LLM-based writing systems increasingly generate "watermarks"---formulaic patterns or predictable structures inherent in AI-generated text---to detect AI-generated contents~\citep{kirchenbauer2024watermarklargelanguagemodels}. 
Without adequate support, 
NNESs may struggle to identify and critically evaluate these watermarks.
Information aids could help users navigate such challenges by visually flagging repetitive structures or suspicious phrases and providing actionable alternatives to improve fluency and originality. 
Such information aids could empower users to critically assess and refine AI-generated content, ensuring their writing remains both original and competitive.

Integrating task-specific information aids into writing systems raises several key research questions. How can these information aids be designed to address the demands of different tasks without overwhelming users? Are there universal aids that remain effective across various writing contexts? 
Addressing these questions could provide valuable insights into the design of adaptable and context-aware writing tools,
ultimately tailored to the diverse needs of NNEss while maintaining usability and minimizing cognitive load.

\subsection{Language Learning Effect of Information Aids}
An intriguing direction for future research is to investigate whether information aids in writing tools can effectively promote language learning. Our study participants expressed optimism about the potential learning benefits of using information aids (\S~\ref{section:findings:impact:4}). Future research could explore this by conducting long-term studies with NNES students who consistently use the system, measuring the learning effects on language proficiency over time. Such research could provide valuable insights into the educational impact of integrating information aids into AI-driven writing systems.

However, findings from related work highlight a potential challenge: using LLMs can sometimes lead to over-reliance on AI, failing to improve users’ language proficiency or even impact cognitive abilities~\citep{zhai2024effects}. This raises an interesting tension between the potential for language learning and the risk of fostering dependence on AI-generated suggestions. Investigating this balance could uncover key design strategies for tools that encourage active learning while minimizing over-reliance.

\subsection{Limitations}
\label{section:discussion:limitations}
Our participant pool consisted exclusively of Korean individuals, which could limit the generalizability of our findings across different linguistic and cultural contexts. Participants were also primarily university students, narrowing the range of age and educational backgrounds represented. Future studies could involve a more diverse sample to examine how language proficiency and cultural variation influence engagement with AI-assisted writing tools.
\rev{This work introduced \system{} as a research artifact and examined its use through a mixed-method in-lab user study, observing real-time interactions and gathering both behavioral logs and reflective feedback in a controlled and task-oriented setting. 
While we sought to create a scenario as realistic as possible within laboratory constraints, in-lab studies may not fully capture the complexity of real-world writing contexts or longer-term usage patterns. Additionally, the scope of statistical analysis was limited by the relatively small number of participants. Future work could incorporate field deployments or longitudinal studies to evaluate system use over time and in more varied scenarios.}
Finally, the fixed interface layout—particularly the consistent ordering of information aid buttons—may have introduced ordering bias that influenced user preferences and interaction patterns. Future iterations could randomize or personalize feature placement to mitigate such effects and more accurately assess user behavior.
\section{Conclusion} 

We investigated how NNESs 
use diverse information aids to assess and select AI paraphrase suggestions.
By developing \system{}, \rev{an AI paraphrasing assistant} designed to integrate multiple information aids and collect user-system interaction data, we observed participants' paraphrasing workflows and preferences with information aids, revealing key patterns in how proficiency levels influence the use of information aids.
While back-translation was the most frequently used aid, it was not sufficient on its own to drive decision-making, highlighting the importance of a combination of aids in supporting informed judgments.
Our findings emphasize the potential of explainable AI paraphrasing tools to empower NNESs by enhancing their confidence, efficiency, and autonomy in writing tasks. However, careful consideration is needed to mitigate the risks of information overload when integrating multiple aids. Building on these insights, we propose actionable design implications for developing AI tools that effectively combine information aids, providing clear, contextualized explanations that meet the diverse needs of NNESs. These findings contribute to advancing explainable AI writing systems for NNESs and open avenues for broader applications in adaptive, user-centered writing support across various domains.

\begin{acks}
We thank our study participants for their valuable feedback on the design of \system{}.
We also thank the DIS 2025 ACs and reviewers for their thoughtful comments and suggestions.
We disclose the use of generative AI tools in the process of writing this manuscript. 
These tools were used solely for editing the authors' own text, and the authors ensured the final content was free from plagiarism, misrepresentation, fabrication, and falsification. 
This work was supported by the National Research Foundation of Korea (NRF) grant funded by the Korea government (MSIT) (RS-2024-00337007). 
\end{acks}

\newpage

\bibliographystyle{ACM-Reference-Format}
\bibliography{_references}

\newpage
\appendix 

\section{CEFR Rubrics}
\label{appendix:formative:cefr}
We list the rubrics we used for describing user English proficiency:

\begin{itemize}[leftmargin=*]
    \item A1 (Beginner): You can understand and use basic phrases and expressions. You can communicate in simple ways when people speak slowly to you.
    \item A2 (Elementary): You can participate in simple exchanges on familiar topics. You can understand and communicate routine information.
    \item B1 (Intermediate): You can communicate in situations and use simple language to communicate feelings, opinions, plans, and experiences.
    \item B2 (Upper Intermediate): You can communicate easily with native English speakers. You can understand and express some complex ideas and topics.
    \item C1 (Advanced): You can understand and use various languages. You can use English flexibly and effectively for social and academic purposes.
    \item C2 (Proficiency): You can understand almost everything you hear or read. You can communicate very fluently and precisely in complex situations.
\end{itemize}

\section{Technical Details}
\label{appendix:technical}

\subsection{Paraphrase Generation}
\label{appendix:technical:paraphrase}

We leveraged GPT-3~\cite{gpt3} to generate multiple paraphrased suggestions. Below is the prompt used for generating paraphrases:

\begin{framed}
    \footnotesize
    \texttt{
    \\
    Generate four paraphrased variations for the given sentence(s) below.\\
    \#\#\#\\
    Sentence(s): ask about\\
    Paraphrased sentence(s):\\
    inquire about\\
    request details on\\
    pose questions about\\
    seek information about\\
    \\
    Sentence(s): Nice to meet you.\\
    Paraphrased sentence(s):\\
    I'm glad to meet you.\\
    Delighted to meet you.\\
    Pleasure to make your acquaintance.\\
    I am pleased to meet you.\\
    \\
    Sentence(s): <ORG\_TXT>\\
    Paraphrased sentence(s):\\
    }
\end{framed}

\noindent The decoding parameters we used are:
\begin{itemize}[leftmargin=*]
    \item \textbf{Engine}: text-davinci-003	
    \item \textbf{Max Tokens}: 250
    \item \textbf{Temperature}: 0.8
    \item \textbf{Top-p}: 0.9
    \item \textbf{Presence Penalty}: 0.5
    \item \textbf{Frequency Penalty}: 0.5
\end{itemize}

\subsection{AI Explanation Generation}
\label{appendix:technical:explanation}
We used GPT-3.5~\cite{gpt3.5turbo} to generate AI Explanation. Below is the prompt we used for generating AI Explanation:

\begin{framed}
    \footnotesize
    \texttt{
    \\
    Compare the paraphrased sentence(s) with the original sentence(s) regarding conveyed tone and appropriate use cases in less than 40 words.
    \\
    \#\#\#\\
    Original: I'm sorry for the late submission of my assignment.\\
    Paraphrased: I apologize for the delay in submitting my assignment.\\
    Comparison: This sentence conveys a more formal tone. It could be used if the audiences are instructors, seniors, or anyone else where formal and respectful language is important.\\
    \\
    Original: Nice to meet you.\\
    Paraphrased: Glad to meet you.\\
    Comparison: This sentence conveys the same tone using different wording. It could be used in either a professional context or a casual social encounter.\\
    \\
    Original: decide to\\
    Paraphrased: resolve to\\
    Comparison: This phrase conveys a more determined and commited tone. It could be used when you want to emphasize your unwavering commitment to a decision.\\
    \\
    Original: I am encountering considerable difficulty in comprehending the course material.\\
    Paraphrased: I'm really struggling to get a good grip on the course material.\\
    Comparison: This sentence conveys a more informal and open tone. It could be used when communicating with friends or peers.\\
    \\
    Original: Could you please provide clarification on the third question in the assignment?\\
    Paraphrased: I was hoping you could help me understand the third question on the assignment.\\
    Comparison: This sentence maintains a polite and inquisitive tone. It could be used when communicating to your instructor or peers in an educational context.\\
    \\
    Original: I'm looking forward to your lecture on Friday.\\
    Paraphrased: I am eagerly anticipating your lecture scheduled for this Friday.\\
    Comparison: This sentence has a more formal tone. It could be used when writing to academics, professionals, or in a formal setting where a higher level of vocabulary is expected.\\
    \\
    Original: <ORG\_TXT>\\
    Paraphrased: <PAR\_TXT>\\
    Comparision:\\
    }
\end{framed}

\noindent The decoding parameters we used are:
\begin{itemize}[leftmargin=*]
    \item \textbf{Engine}: gpt-3.5-turbo	
    \item \textbf{Max Tokens}: inf
    \item \textbf{Temperature}: 1
    \item \textbf{Top-p}: 1
    \item \textbf{Presence Penalty}: 0
    \item \textbf{Frequency Penalty}: 0
\end{itemize}

\section{User Study}
\label{appendix: study}

\subsection{Email Scenarios in the User Study}
\label{appendix:main:emailscenarios}

% table
\begin{table*}[t]
\centering\small
\renewcommand{\arraystretch}{1.1}
% \resizebox{\linewidth}{!}{
\begin{tabular}[t]{lll} %{@{}lcllll@{}}
\toprule
 & \textbf{Email Scenario} \\ \midrule
1 &
  \begin{tabular}[c]{@{}l@{}}Having attended a lecture on topic X, you are intrigued and wish to explore it further. You email the professor, seeking \\recommendations for related papers to expand your understanding. In your email, politely express your appreciation \\for the lecture, convey your enthusiasm for the subject, and request recommendations for relevant academic papers.\end{tabular} \\ \midrule
2 &
  \begin{tabular}[c]{@{}l@{}}After reviewing your graded assignment, you notice an error in the calculation of your final score. You write an email \\to your professor explaining the issue to request a grade correction. In your email, politely explain the situation and \\express your concern about the potential impact on your grade. Show appreciation for their attention to the matter and\\ your hope for a prompt resolution.\end{tabular} \\ \midrule
3 &
  \begin{tabular}[c]{@{}l@{}}You have written a personal statement for your internship application. To refine your personal statement further, you've \\decided to reach out to your English professor and ask for their feedback to improve it. In the email, politely explain your \\purpose for sending the email, ask for their constructive feedback on the personal statement you attached, and appreciate \\their help.\end{tabular} \\ \midrule
4 &
  \begin{tabular}[c]{@{}l@{}}You missed the class because you were sick and had to go to the hospital. You email your professor to ask if you can get an \\excused absence by presenting a medical record. When writing an email, politely explain your situation and show \\appreciation for your professor's understanding and consideration.\end{tabular} \\ \midrule
5 &
  \begin{tabular}[c]{@{}l@{}}You are interested in auditing the course X at your university. You email the professor responsible for teaching course X\\ to inquire about the possibility of auditing their course. In your email, briefly introduce yourself and politely explain your \\intention to audit the course. Inquire about the professor's policy on auditing and convey your enthusiasm for the subject.\end{tabular} \\ \midrule
6 &
  \begin{tabular}[c]{@{}l@{}}You are planning to study abroad as an exchange student in the upcoming semester. However, you missed the deadline for \\the dormitory application, resulting in no dormitory assignment. You want to reach out to the university to inquire if they\\allow for a late dormitory application. In the email, politely express your situation, apologize for the oversight, and ask if \\there is a solution.\end{tabular} \\ 
\bottomrule
\end{tabular}
% }
\vspace{1em}
\caption{The six email scenarios we used in the Main Task. We created the scenarios from the email excerpts from the recruitment survey.}
\Description{The table lists the six email scenarios used in the main study.}
\label{tab:main:emailscenarios}
\end{table*}
Table~\ref{tab:main:emailscenarios} shows the email scenarios we provided in the main study. The scenarios were created from the samples of the emails we received from the screening survey.

\subsection{Thematic Analysis Codebook}
\label{appendix:study:codebook}
\rev{Table~\ref{tab:appendix:codebook} and Table~\ref{tab:appendix:codebook2} shows the codebook used for the thematic analysis of user experience on \system{}.}

\begingroup
\setlength{\tabcolsep}{8pt} % Default value: 6pt 
\begin{table*}[t]

\caption{\rev{Codebook summarizing the dimensions, codes, code descriptions, and example quotes from participants regarding the usage purpose and patterns of \system{}}}
\Description{This codebook presents the dimensions, codes, code descriptions, and example quotes from participants regarding the usage purpose and patterns of \system{}.}
\label{tab:appendix:codebook}

\centering
\renewcommand{\arraystretch}{1.25} 
\scalebox{0.9}{ % adjust the scale factor as needed
\rev{ % Change font color to blue
\begin{tabular}{p{4cm}p{5cm}p{7cm}}
\toprule
 \textbf{Dimensions and Codes} & \textbf{Code Description} & \textbf{Example Quote} \\ 
    \midrule
    \multicolumn{2}{l}{\textit{What are users' purposes of using information aids?}} \\
    \dashedline{1-3}

    Preserve intended meaning & Ensure the suggestion retains the original meaning. & ``\textit{I checked whether the suggested sentence included everything necessary or if anything important was missing or distorted.}'' \\ 
    
    Assess tone appropriateness & Ensure tone and nuance fit the context and audience. & ``\textit{I mainly used it to see if what I wrote was appropriate for the situation and the intended reader.}'' \\

    Explore real-world usage & Examine how expressions are used in real-world contexts. & 
    ``\textit{I used it to see how unfamiliar expressions are used in actual sentences.}'' \\

 Verify common usage & Confirm whether a phrase is commonly used or outdated. & ``\textit{It looked natural to me, but I wanted to make sure it was something people commonly say nowadays---like checking if it’s not an outdated phrase no longer in use.}'' \\
    
    \midrule
    \multicolumn{2}{l}{\textit{How do users use information aids to evaluate suggestions?}} \\
    \dashedline{1-3}

    Screen for best fit & Narrow down options by dismissing weak suggestions. & ``\textit{My general process was to first filter out bad ones, then pick the best among the remaining options.}'' \\

    Validate initial preference & Use features to validate an initial judgment. & ``\textit{Among the four options, I initially liked one the most and thought I would go with it. As I explored the features one by one, they confirmed what I had sensed---that it was the most polite and formal, just as I had thought.}'' \\

    Cross-check information aids & Use one feature to interpret or validate another. & ``\textit{When I only looked at the AI Score, I doubted its reliability and couldn’t understand why it rated a suggestion the highest. But when I checked the AI Translation and Explanation, it made sense, and I could accept the reasoning behind the score.}'' \\

    Break tie with information aids & Let features guide selection when unsure between options. & ``\textit{If I was still unsure which of two sentences to use, I thought, perhaps the one with the higher numerical value would be better.}'' \\
    
    \bottomrule
\end{tabular}
} % end text color
}
\vspace{1em}
\end{table*}
\renewcommand{\arraystretch}{1} 
\endgroup
\begingroup
\setlength{\tabcolsep}{8pt} % Default value: 6pt 
\begin{table*}[t]

\caption{\rev{Codebook summarizing the dimensions, codes, code descriptions, and example quotes from participants regarding the user experiences and suggestions for \system{}}}
\Description{This codebook presents the dimensions, codes, code descriptions, and example quotes from participants regarding the user experience and suggestions for \system{}.}
\label{tab:appendix:codebook2}

\centering
\renewcommand{\arraystretch}{1.25} 
\scalebox{0.9}{ % adjust the scale factor as needed
\rev{ % Change font color to blue
\begin{tabular}{p{4cm}p{5cm}p{7cm}}
\toprule
 \textbf{Dimensions and Codes} & \textbf{Code Description} & \textbf{Example Quote} \\ 
    \midrule
    \multicolumn{2}{l}{\textit{What are perceived impacts of information aids on user experience?}} \\
    \dashedline{1-3}

    Informed decision-making & Having access to diverse information aids supported more confident and informed choices. & ``\textit{Having access to various types of information that either confirmed or refuted the suitability of a suggestion made me feel more reassured and confident in my choice.}'' \\

    Improved efficiency & Having diverse information aids integrated in one interface streamlined the suggestion selection process. & ``\textit{Typically, my writing process involves constantly switching between windows: composing, searching dictionaries, and consulting ChatGPT. This system consolidates these actions within one platform, significantly aiding my workflow.}''\\

    Cognitive overload & The abundance of information aids sometimes caused mental fatigue and inefficiency. & ``\textit{I feel compelled to use every information aid, even for sentences I could easily move on from. I don’t think this process is efficient, but I am nonetheless convinced that it improves writing.}''\\

    Increased autonomy & Interacting with information aids enhanced agency and authorship in the writing process. & ``\textit{Utilizing features to understand how suggestions might sound and integrating these suggestions into my text, made me feel more actively involved in the writing process. It felt as if I were crafting the text myself.}''\\

    Potential language learning & Participants perceived interacting with information aids as opportunities for language learning. & ``\textit{Example Sentence, unlike other features that provide direct hints, allows for the indirect exploration of various sentences and fosters thoughtful consideration.}''\\

    \midrule
    \multicolumn{2}{l}{\textit{What are the factors that influence users' feature usage experience?}} \\
    \dashedline{1-3}

    Simplicity of presentation & Participants preferred concise, direct information presentation format. & ``\textit{I needed to make quick decisions on suggestions, but since it’s presented in long paragraphs, it wasn’t easy to skim. If it were in bullet points, it would have been faster and more convenient.}'' \\

    Explainability & Participants expressed a need for explainability and transparency about how numeric features were derived. & ``\textit{It wasn’t clear what database the trends shown in Frequency were based on. For example, is this frequency higher because the word appears often in news articles? Knowing this would make the feature more helpful.}'' \\

    Personalization & Participants wished for an interface tailored to personalized usage patterns. & ``\textit{When using a tool repeatedly, people tend to stick to certain features. Instead of showing all features, it might be better to display only the ones I actively use.}'' \\
    
    Interactive refinement of suggestions & Participants wished to directly modify suggestions without leaving the interface or fully rewriting the text. & ``\textit{I didn’t want to completely rewrite my sentence; I aimed to keep the original structure intact while making minor adjustments to the words or expressions.}'' \\
    
    Incorporate original text for comparison & Participants found it difficult to evaluate suggestions without seeing their original input. & ``\textit{It would be more convenient if my original input was displayed alongside the suggestions for direct comparison. Paraphrasing provides four candidates, but the original sentence is essentially a fifth option. Including translations or scores for the original would make it easier to evaluate all options equally.}'' \\

    \bottomrule
\end{tabular}
} % end text color
}
\vspace{1em}
\end{table*}
\renewcommand{\arraystretch}{1} 
\endgroup

\end{document}